\documentclass[a4paper,11pt]{article}
\usepackage{amsfonts, amsmath,amssymb,bm,enumerate, graphicx,color}
\usepackage{subfigure}
\usepackage{epstopdf}
\usepackage[dvipsnames]{xcolor}
\usepackage[colorlinks=true,urlcolor=blue,citecolor=purple,linkcolor=blue,bookmarks=true]{hyperref}
\newtheorem{The}{Theorem}
\newtheorem{Lem}[The]{Lemma}
\newtheorem{Cor}[The]{Corollary}
\newtheorem{Def}[The]{Definition}
\newtheorem{Exa}[The]{Example}

\numberwithin{The}{section}
\allowdisplaybreaks

\newcommand{\dif}{\mathrm{d}}
\def\Fbar{ {\overline F}}
\def\qed{\hfill \vrule height 6pt width 6pt depth 0pt}
\def\proof{\trivlist \item[\hskip \labelsep{\bf Proof.\ }]}

\newcounter{quest}

\begin{document}
\title{\bf Allocations of Cold Standbys to Series and Parallel Systems with Dependent Components\thanks{Corresponding author: Yiying Zhang. E-mail: yyzhang@nankai.edu.cn}}

\author{Xiaoyu Zhang\\[5pt]
School of Mathematics and Statistics\\
Jiangsu Normal University, Xuzhou 221116, P. R. China\\[10pt]
Yiying Zhang\\[5pt]
School of Statistics and Data Science, LPMC and KLMDASR\\
Nankai University, Tianjin 300071, P. R. China\\
{\sf E-mail: yyzhang@nankai.edu.cn}\\[10pt]
Rui Fang\\[5pt]
Department of Mathematics\\
Shantou University, Shantou 515063, P. R. China\\[10pt]}

\maketitle

\begin{abstract}
In the context of industrial engineering, cold-standby redundancies allocation strategy is usually adopted to improve the reliability of coherent systems. This paper investigates optimal allocation strategies of {cold standbys} for {series and parallel} systems comprised of dependent components with left/right tail weakly stochastic arrangement increasing lifetimes. For the case of heterogeneous and independent matched {cold standbys}, it is proved that better redundancies should be put in the nodes having weaker [better] components for series [parallel] systems. For the case of homogeneous and independent {cold standbys}, it is shown that more redundancies should be put in standby with weaker [better] components to enhance the reliability of series [parallel] systems. The results developed here generalize and extend those corresponding ones in the literature to the case of series and parallel systems with dependent components. Numerical examples are also presented to provide guidance for the practical use of our theoretical findings.\\


\medskip


\end{abstract}

\section{Introduction}
In reliability engineering and system security, one common way to optimize system performance is to introduce  redundancies (or spares) to the components. Three types of redundancies are usually used to improve the performance of reliability systems, i.e., the hot standby, the cold standby, and the warm standby. For the hot standby, available spares are put in parallel with the original components and function simultaneously with them. For the cold standby, concerned spares are attached to the components of the system in such a way that the redundancies start to work immediately after the failures of the original components. The warm standby is a redundancy type between the hot standby and the cold standby. For the case of warm standby, the redundant components may fail to work even before its operation; indeed, the warm redundancies are more likely to be damaged than the cold standbys. Moreover, the failure rate of an inactive warm standby is less than its actual failure rate, and then it is switched to the active state immediately after the failure of the original component and its failure rate will increase. Interested readers may refer to Yun and Cha \cite{yun10}, Hazra and Nanda \cite{hazra15}, Finkelstein et al. \cite{Finkelstein18}, and Hadipour et al. \cite{Hadipour18} for more detailed discussions. For all of these three types of standbys, the system performance under different allocation policies can be effectively evaluated via stochastic comparisons in terms of various stochastic orders.


A system is said to be coherent if each component is relevant and the structure function is increasing in its components (c.f. Barlow and Proschan, \cite{bp81}). The past several decades have witnessed comprehensive developments on investigating optimal allocation policies of hot standbys (i.e., active redundancies) for coherent systems (especially $k$-out-of-$n$ systems) consisting of independent components; see for example Boland et al. \cite{boland88,boland92}, Singh and Misra \cite{singh94}, Vald\'{e}s and Zequeira \cite{valdes06}, Vald\'{e}s et al. \cite{valdes10}, Brito et al. \cite{brito11}, Misra et al. \cite{misra11a}, Hazra and Nanda \cite{hazra14}, Zhao et al. \cite{zhaonrl13,zhaoorl13,zhaoejor15,zhaoejor17}, Da and Ding \cite{Da16},  Ding et al. \cite{ding17}, and Zhang \cite{zhang18}. On the other hand, some research work has appeared on redundancies allocation for coherent systems with dependent components. For instance, Belzunce et al. \cite{belzunce11} and Belzunce et al. \cite{belzunce13} considered {hot standbys}  allocation for $k$-out-of-$n$ systems comprised of statistically dependent components with their lifetimes characterized by joint stochastic orders (c.f. Shanthikumar and Yao, \cite{shan91}). Interested readers may refer to You and Li \cite{youli14}, You et al. \cite{you16}, and Zhang et al. \cite{asmbi17} for more study along this direction.



On account of the complexity of distribution theory, there is not much work on studying optimal allocation strategies of {cold-standby} redundancies. Boland et al. \cite{boland92} might be the first to investigate how to optimally assign {cold-standby} redundancies to series and parallel systems. They showed that the optimal allocation strategy for a series system is opposite to that for a parallel system when the original components and the redundancies are i.i.d. After that, many researchers have paid attention to the allocation problem of {cold-standby} redundancies in series and parallel systems; see for instance Singh and Misra \cite{singh94}, Li et al. \cite{liyanhu11}, Misra et al. \cite{misra11b}, Zhuang and Li \cite{zhuang15}, Doostparast \cite{Doost17}, and Chen et al. \cite{chen18}. Another research stream focuses on the effects of {cold-standby} redundancies on the performance of coherent systems. For more details, readers are referred to da Costa Bueno \cite{daCosta05}, Ardakan and Hamadani \cite{ardakan14}, Eryilmaz \cite{erylimaz17}, and Gholinezhad and Hamadani \cite{gho17}, Eryilmaz and Erkan \cite{erylimaz18}, and the references therein.

To the best of our knowledge, {rare work exists on studying optimal allocation of {cold-standby} redundancies for systems with interdependent components except Belzunce et al. \cite{belzunce13} and Jeddi and Doostparast \cite{jeddi16}}. Belzunce et al. \cite{belzunce13} established the optimal allocation policy for one cold-standby redundancy in series and parallel systems by means of the stochastic precedence order and the usual stochastic order when the lifetimes of the original components are ordered via joint stochastic orders. Jeddi and Doostparast \cite{jeddi16} studied the same  allocation problem by employing quadratic dependence orderings (see Shaked and Shanthikumar, \cite{ss07}). However, none of these results treats the case of more than two cold-standby redundancies. The objective of the present paper is to fill this gap through pinpointing optimal allocation strategies of cold-standby redundancies for series and parallel systems with heterogeneous and dependent components.

It should be remarked that our method is quite different with those of Belzunce et al. \cite{belzunce13} and Jeddi and Doostparast \cite{jeddi16}. At the preliminary  working stage of a system with {cold-standby} redundancies, it is reasonable to assume that the original components are positively interdependent (due to the external stress or shock and common environment), while the {cold-standby} redundancies are assumed to be independent since they are not activated before the failures of original components. In this paper, we assume the components of series/parallel systems are positively dependent and have left tail weakly stochastic arrangement increasing (LWSAI) or right tail weakly stochastic arrangement increasing (RWSAI) lifetimes. The concerned {cold-standby} spares are assumed to be statistically independent and they are also independent of the original components. Several stochastic orders including the usual stochastic order, the increasing convex order, and the increasing concave order are employed to derive the optimal allocation policies. More explicitly, for the case of heterogeneous and independent matched {cold-standby} redundancies, we prove that the better redundancy should be put in the node with weaker [better] component for a series [parallel] system. For the case of homogeneous and independent {cold-standby} redundancies, we show  that more redundancies should be allocated to the weaker [better] component to enhance the reliability of a series [parallel] system. The results developed here generalize and extend those related ones in Singh and Misra \cite{singh94}, Misra et al. \cite{misra11b}, Belzunce et al. \cite{belzunce13}, and Jeddi and Doostparast \cite{jeddi16}.

The remainder of the paper is rolled out as follows. Section \ref{preli} recalls some pertinent notions and definitions used in the sequel. In Section \ref{sechetero}, optimal allocation strategies of heterogeneous and independent matched {cold-standby} redundancies are presented for series and parallel systems comprised of LWSAI or RWSAI components. In Section \ref{sechomo}, optimal allocations are investigated for the case of a batch of i.i.d. {cold-standby} redundancies in series and parallel systems. Section \ref{seccon} concludes the paper.

\section{Preliminaries}\label{preli}
Throughout, the terms \emph{increasing} and \emph{decreasing} are used in a non-restrict sense. Let $\mathbb{R}=(-\infty,+\infty)$, $\mathbb{R}_{+}=[0,+\infty)$, and $\mathbb{N}=\{0,1,2,3,\ldots\}$. All random  variables are assumed to be non-negative, and all expectations are well defined whenever they appear.

\begin{Def} For any two non-negative random variables $X$ and $Y$, let $f_{X}$ and $f_{Y}$, $F_{X}$ and $F_{Y}$, $\overline{F}_{X}$ and $\overline{F}_{Y}$ be their density, distribution, and survival functions, respectively. Then, $X$ is said to be smaller than $Y$ in the
\begin{enumerate}
\item [{(i)}] {likelihood ratio order} (denoted by $X \leq_{\rm lr} Y$) if $f_{Y}(x)/f_{X}(x)$ is increasing in $x\in\mathbb{R}_{+}$;
\item [{(ii)}] {hazard rate order} (denoted by $X \leq_{\rm hr} Y$) if $\overline{F}_{Y}(x)/\overline{F}_{X}(x)$ is increasing in $x\in\mathbb{R}_{+}$;
\item [{(iii)}] {reversed hazard rate order} (denoted by $X \leq_{\rm rh} Y$) if $F_{Y}(x)/F_{X}(x)$ is increasing in $x\in\mathbb{R}_{+}$;
\item [{(iv)}] {usual stochastic order} (denoted by $X \leq_{\rm st} Y$) if $\overline{F}_{X}(x)\leq\overline{F}_{Y}(x)$ for all $x\in\mathbb{R}_{+}$;
\item [{(v)}] {increasing convex order} (denoted by $X \leq_{\rm icx} Y$) if $\mathbb{E}[\phi(X)]\leq \mathbb{E}[\phi(Y)]$ for any increasing and convex function $\phi$;
\item [{(vi)}] {increasing concave order} (denoted by $X \leq_{\rm icv} Y$) if $\mathbb{E}[\phi(X)]\leq \mathbb{E}[\phi(Y)]$ for any increasing and concave function $\phi$.
\end{enumerate}
\end{Def}

It is well known that
\begin{equation*}
X \leq_{\rm lr} Y\Longrightarrow X \leq_{\rm hr ~[or~rh]} Y\Longrightarrow X \leq_{\rm st} Y\Longrightarrow X \leq_{\rm icx~[or~icv]} Y.
\end{equation*}
One may refer to Shaked and Shanthikumar \cite{ss07} for comprehensive discussions on the properties and applications of above mentioned stochastic orders.

\medskip

Next, we recall several useful dependence notions of arrangement increasing (AI). For any function $g:\mathbb{R}^{n}\mapsto\mathbb{R}$ and any pair $(i,j)$ such that $1\leq i<j\leq n$, let
\begin{equation*}
\mathcal{G}_{s}^{i,j}(n)=\{g(\bm{x}): g(\bm{x})\geq g(\tau_{i,j}(\bm{x}))~\mbox{for any $x_{i}\leq x_{j}$}\},
\end{equation*}
\begin{equation*}
\mathcal{G}_{l}^{i,j}(n)=\{g(\bm{x}): g(\bm{x})-g(\tau_{i,j}(\bm{x}))~\mbox{is decreasing in $x_{i}\leq x_{j}$}\},
\end{equation*}
\begin{equation*}
\mbox{and}\quad\mathcal{G}_{r}^{i,j}(n)=\{g(\bm{x}): g(\bm{x})-g(\tau_{i,j}(\bm{x}))~\mbox{is increasing in $x_{j}\geq x_{i}$}\},
\end{equation*}
where $\tau_{i,j}(\bm{x})$ denotes the permutation of $\bm{x}$ with its $i$-th and $j$-th components exchanged.

\begin{Def} A random vector $\bm{X}=(X_{1},\ldots,X_{n})$ is said to be
\begin{itemize}
\item [(i)] {stochastic arrangement increasing} (SAI) if $\mathbb{E}[g(\bm{X})]\geq\mathbb{E}[g(\tau_{i,j}(\bm{X}))]$ for any $g\in\mathcal{G}_{s}^{i,j}(n)$ and all pair $(i,j)$ such that $1\leq i<j\leq n$;
\item [(ii)] {left tail weakly stochastic arrangement increasing} (LWSAI) if $\mathbb{E}[g(\bm{X})]\geq\mathbb{E}[g(\tau_{i,j}(\bm{X}))]$ for any $g\in\mathcal{G}_{l}^{i,j}(n)$ and all pair $(i,j)$ such that $1\leq i<j\leq n$;
\item [(iii)] {right tail weakly stochastic arrangement increasing} (RWSAI) if $\mathbb{E}[g(\bm{X})]\geq\mathbb{E}[g(\tau_{i,j}(\bm{X}))]$ for any $g\in\mathcal{G}_{r}^{i,j}(n)$ and all pair $(i,j)$ such that $1\leq i<j\leq n$.
\end{itemize}
\end{Def}

Since {$\mathcal{G}_{s}^{i,j}(n)\supset\mathcal{G}_{l}^{i,j}(n)~[\mathcal{G}_{r}^{i,j}(n)]$}, SAI implies both LWSAI and RWSAI. Multivariate versions of Dirichlet distribution, inverted Dirichlet distribution, $F$ distribution, and Pareto distribution of type I (see Hollander et al., \cite{hollander77}) are all SAI and hence are RWSAI and LWSAI whenever the corresponding parameters are arrayed in the ascending order. SAI, LWSAI, and RWSAI were proposed by Cai and Wei \cite{caiwei14,caiwei15} and have been applied in the fields of financial engineering and actuarial science; see for example Cai and Wei \cite{caiwei15}, Zhang and Zhao \cite{zhang15}, You and Li \cite{youli15}, and Zhang et al. \cite{astin18}. According to Cai and Wei \cite{caiwei14,caiwei15}, the following chain of implications always holds:
\begin{equation}\label{saiimplication}
{\rm SAI}\Longrightarrow{\rm LWSAI}~{\rm[RWSAI]}\Longrightarrow X_{1}^{\perp}\leq_{\rm st}\cdots \leq_{\rm st}X_{n}^{\perp},
\end{equation}
where $X_{1}^{\perp},\ldots,X_{n}^{\perp}$ are the independent version of $\bm{X}$. In this paper, we shall employ these useful notions to characterize the dependence structure of components lifetimes in series and parallel systems.

The \emph{joint multivariate likelihood ratio order} and the \emph{joint multivariate reversed hazard rate order}, which were introduced by Shanthikumar and Yao \cite{shan91}, compare random variables by taking into account the statistical dependence. These two types of joint stochastic orders are equivalent to SAI and LWSAI, respectively. {Therefore, the results established in Theorems \ref{heterseries} and \ref{heterparallel} cover Theorem 3.3(b) and Theorem 3.5 of Belzunce et al. \cite{belzunce13}, respectively.} Also, for absolutely continuous random vectors, LWSAI coincides with LTPD according to Proposition 3.7 of Cai and Wei \cite{caiwei15}. For comprehensive treatments on other interesting higher joint stochastic orders, please refer to Wei \cite{wei17}.

\medskip

For a random vector $\bm{X}=(X_1,\ldots,X_n)$ with joint distribution function $H$ and univariate marginal distribution functions $F_1,\ldots,F_n$, its \emph{copula} is a distribution function $C:[0,1]^n\mapsto [0,1]$, satisfying
\begin{equation*}
  H(\bm{x})=C(F_1(x_1),\ldots,F_{n}(x_n)).
\end{equation*}
Similarly, a \emph{survival copula} is a distribution function $\overline{C}:[0,1]^n\mapsto [0,1]$, satisfying
\begin{equation*}
  \overline{H}(\bm{x})=\mathbb{P}(X_1>x_1,\ldots,X_n>x_n)=\overline{C}(\Fbar_1(x_1),\ldots,\Fbar_n(x_n)),
\end{equation*}
where $\Fbar_i=1-F_i$, for $i=1,\ldots,n$, are marginal survival functions and $\overline{H}(\bm{x})$ is the joint survival function. The copula does not include any information of marginal distributions, and thus it provides us a particularly convenient way to impose a dependence structure on predetermined marginal distributions in practice. Archimedean copulas are rather popular due to its mathematical tractability and the capability of capturing wide ranges of dependence. By definition, for a decreasing and continuous function $\phi:[0,+\infty)\mapsto[0,1]$ such that $\phi(0)=1$ and $\phi(+\infty)=0$,
\begin{equation*}
C_{\phi}(u_{1},\cdots,u_{n})=\phi\left(\sum_{i=1}^n\phi^{-1}(u_{i})\right),\quad\mbox{ for $u_{i}\in[0,1]$, $i=1,2,\ldots,n,$}
\end{equation*}
is called an \emph{Archimedean copula} with the generator $\phi$ if $(-1)^{k}\phi^{(k)}(x)\geq0$ for $k=0,\ldots,n-2$ and $(-1)^{n-2}\phi^{(n-2)}(x)$ is decreasing and convex. The Archimedean family contains many well-known copulas, including the independence (or product) copula, the Clayton copula, and the Ali-Mikhail-Haq (AMH) copula. For more detailed study  on the properties of copulas, one may refer to Nelsen \cite{nelsen06}.

In accordance with Theorem 5.7 of Cai and Wei \cite{caiwei14}, $\bm{X}$ is RWSAI if $X_{1}\leq_{\rm hr}\cdots\leq_{\rm hr}X_{n}$ and are connected with an Archimedean survival copula with log-convex generator. Proposition 4.1 of Cai and Wei \cite{caiwei15} shows that $\bm{X}$ is LWSAI if $X_{1}\leq_{\rm rh}\cdots\leq_{\rm rh}X_{n}$ and share an Archimedean copula with log-convex generator. According to Corollary 8.23(b) of Joe \cite{joe14}, $\bm{X}$ is \emph{positive lower orthant dependent} (PLOD) if the generator is log-convex, which indicates some kind of positive dependence structure.

\section{Heterogeneous cold-standby redundancies}\label{sechetero}
For two non-negative $n$-dimensional real vectors $\bm{x}$ and $\bm{y}$, we denote $\min(\bm{x}+\bm{y})=\min(x_{1}+y_{1},\ldots,x_{n}+y_{n})$, $\max(\bm{x}+\bm{y})=\max(x_{1}+y_{1},\ldots,x_{n}+y_{n})$, and the sub-vector
\begin{equation*}
{\bm x}_{\{i,j\}}=(x_{1},\ldots,x_{i-1},x_{i+1},\ldots,x_{j-1},x_{j+1},\ldots,x_{n}),~\mbox{for $1\leq i<j\leq n$.}
\end{equation*}
Consider a system consisting of heterogeneous components $C_{1},\ldots,C_{n}$ with $C_{i}$ having lifetime $X_{i}$, for $i=1,\ldots,n$. Let $Y_{i}$ be the lifetime of cold-standby redundancy $R_{i}$ allocated to component $C_{i}$, for $i=1,\ldots,n$. Then, the lifetimes of the resulting series and parallel systems can be denoted by $\min(\bm{X}+\bm{Y})$ and $\max(\bm{X}+\bm{Y})$, respectively. Hereafter, it is assumed that $X_{i}$'s are dependent through SAI, LWSAI or RWSAI, while the lifetimes $Y_{i}$'s are assumed to be independent and they are also independent of $X_{i}$'s.

\subsection{Series system}
This subsection deals with optimal allocations of $n$ heterogeneous and independent redundancies to a series system consisting of $n$ dependent components. To begin with, let us introduce a useful lemma presenting functional characterizations of LWSAI and RWSAI bivariate random vectors.

\begin{Lem}{\rm (You and Li \cite{youli15})}\label{bivachar} A bivariate random vector $(X_{1},X_{2})$ is LWSAI [RWSAI] if and only if $\mathbb{E}[g_{2}(X_{1},X_{2})]\geq\mathbb{E}[g_{1}(X_{1},X_{2})]$ for all $g_{1}$ and $g_{2}$ such that
\begin{itemize}
\item [(i)] $g_{2}(x_{1},x_{2})-g_{1}(x_{1},x_{2})$ is decreasing [increasing] in $x_{1}\leq x_{2}$ [$x_{2}\geq x_{1}$] for any $x_{2}$ [$x_{1}$];
\item [(ii)] $g_{2}(x_{1},x_{2})+g_{2}(x_{2},x_{1})\geq g_{1}(x_{1},x_{2})+g_{1}(x_{2},x_{1})$ for any $x_{2}\geq x_{1}$.
\end{itemize}
\end{Lem}

Now, we present the first main result.
\begin{The}\label{heterseries} Suppose that $Y_{1}\geq_{\rm lr}Y_{2}\geq_{\rm lr}\cdots\geq_{\rm lr}Y_{n}$. For any $1\leq i<j\leq n$,
\begin{itemize}
\item [(i)]if $\bm{X}$ is LWSAI, then $\min(\bm{X}+\bm{Y})\geq_{\rm icv}\min(\bm{X}+\tau_{i,j}(\bm{Y}))$;
\item [(ii)]if $\bm{X}$ is RWSAI, then $\min(\bm{X}+\bm{Y})\geq_{\rm st}\min(\bm{X}+\tau_{i,j}(\bm{Y}))$.
\end{itemize}
\end{The}
\proof Let $p_{i}(\cdot)$ be the density function of $Y_{i}$, for $i=1,\ldots,n$. For any $1\leq i<j\leq n$ and any integrable function $u$, it is easy to verify that
\begin{eqnarray}\label{rwequa1}
\lefteqn{\mathbb{E}[u(\min(\bm{X}+\bm{Y}))]-\mathbb{E}[u(\min(\bm{X}+\tau_{i,j}(\bm{Y})))]}&& \nonumber\\
  &=& \idotsint\limits_{\mathbb{R}_{+}^{n-2}}\prod_{k\neq i,j}^{n}p_{k}(y_{k})\dif y_{k}\iint\limits_{y_{i}\leq y_{j}}\mathbb{E}[u(\min(\bm{X}+\bm{y}))]p_{i}(y_{i})p_{j}(y_{j})\dif y_{i}\dif y_{j}\nonumber\\
  &&+\idotsint\limits_{\mathbb{R}_{+}^{n-2}}\prod_{k\neq i,j}^{n}p_{k}(y_{k})\dif y_{k}\iint\limits_{y_{i}\leq y_{j}}\mathbb{E}[u(\min(\bm{X}+\tau_{i,j}(\bm{y})))]p_{i}(y_{j})p_{j}(y_{i})\dif y_{i}\dif y_{j}\nonumber\\
  &&-\idotsint\limits_{\mathbb{R}_{+}^{n-2}}\prod_{k\neq i,j}^{n}p_{k}(y_{k})\dif y_{k}\iint\limits_{y_{i}\leq y_{j}}\mathbb{E}[u(\min(\bm{X}+\tau_{i,j}(\bm{y})))]p_{i}(y_{i})p_{j}(y_{j})\dif y_{i}\dif y_{j}\nonumber\\
  & &- \idotsint\limits_{\mathbb{R}_{+}^{n-2}}\prod_{k\neq i,j}^{n}p_{k}(y_{k})\dif y_{k}\iint\limits_{y_{i}\leq y_{j}}\mathbb{E}[u(\min(\bm{X}+\bm{y}))]p_{i}(y_{j})p_{j}(y_{i})\dif y_{i}\dif y_{j} \nonumber\\
  &=&\idotsint\limits_{\mathbb{R}_{+}^{n-2}}\prod_{k\neq i,j}^{n}p_{k}(y_{k}) \dif y_{k}\iint\limits_{y_{i}\leq y_{j}}[\mathbb{E}[u(\min(\bm{X}+\bm{y}))]-\mathbb{E}[u(\min(\bm{X}+\tau_{i,j}(\bm{y})))]]\nonumber\\
  &&\quad\times[p_{i}(y_{i})p_{j}(y_{j})-p_{i}(y_{j})p_{j}(y_{i})]\dif y_{i}\dif y_{j}.
\end{eqnarray}
In the next, we prove that the integrand of (\ref{rwequa1}) is non-negative. First, upon using the condition $Y_{i}\geq_{\rm lr}Y_{j}$ for $i<j$, we know that $p_{i}(y)/p_{j}(y)$ is increasing in $y\in\mathbb{R}_{+}$. Thus, it must hold that
\begin{equation}\label{densieq1}
 p_{i}(y_{i})p_{j}(y_{j})\leq p_{i}(y_{j})p_{j}(y_{i}), \quad y_{i}\leq y_{j}.
\end{equation}
Now, it suffices to show that
\begin{equation}\label{condieq2}
\mathbb{E}[u(\min(\bm{X}+\bm{y}))]\leq\mathbb{E}[u(\min(\bm{X}+\tau_{i,j}(\bm{y})))],\quad y_{i}\leq y_{j},
\end{equation}
for increasing concave $u$ under (i) and increasing $u$ under (ii).

\underline{Proof of (i): $\bm{X}$ is LWSAI.} In light of Proposition 3.2 of Cai and Wei \cite{caiwei15}, the LWSAI property of $\bm{X}$ implies that $[(X_{i},X_{j})|\bm{X}_{\{i,j\}}]$ is LWSAI. For any given $\bm{X}_{\{i,j\}}=\bm{x}_{\{i,j\}}$, let
\begin{equation*}
g_{1}(x_{i},x_{j})=u(\min(\bm{x}+\bm{y}))=u(\min\{x_{i}+y_{i},x_{j}+y_{j},\min\{(\bm{x}+\bm{y})_{\{i,j\}}\}\})
\end{equation*}
and
\begin{equation*}
g_{2}(x_{i},x_{j})=u(\min(\bm{x}+\tau_{i,j}(\bm{y})))=u(\min\{x_{i}+y_{j},x_{j}+y_{i},\min\{(\bm{x}+\bm{y})_{\{i,j\}}\}\}).
\end{equation*}
Thus, for any $y_{i}\leq y_{j}$ and $x_{i}\leq x_{j}$ we have
\begin{eqnarray}\label{deltaeq1}
\Delta_{1}(x_{i},x_{j})&=&g_{2}(x_{i},x_{j})-g_{1}(x_{i},x_{j})\nonumber\\
&=& u(\min\{x_{i}+y_{j},x_{j}+y_{i},\min\{(\bm{x}+\bm{y})_{\{i,j\}}\}\})\nonumber\\
&&\quad-u(\min\{x_{i}+y_{i},\min\{(\bm{x}+\bm{y})_{\{i,j\}}\}\}).
\end{eqnarray}
On the one hand, it can be checked that, for any $x_{j}\geq x_{i}$,
\begin{equation}\label{lemcon2}
g_{2}(x_{i},x_{j})+g_{2}(x_{j},x_{i})=g_{1}(x_{i},x_{j})+g_{1}(x_{j},x_{i}).
\end{equation}
On the other hand, in order to prove that $g_{2}(x_{i},x_{j})-g_{1}(x_{i},x_{j})$ is decreasing in $x_{i}\leq x_{j}$ for any increasing concave $u$, the following several cases are considered.

\underline{Case 1: $x_{i}+y_{j}\geq x_{j}+y_{i}$.} For this case, it is clear that
\begin{equation*}
\Delta_{1}(x_{i},x_{j})=u(\min\{x_{j}+y_{i},\min\{(\bm{x}+\bm{y})_{\{i,j\}}\}\})-u(\min\{x_{i}+y_{i},\min\{(\bm{x}+\bm{y})_{\{i,j\}}\}\})
\end{equation*}
is decreasing in $x_{i}\leq x_{j}$ for any $x_{j}$ and any increasing concave $u$.

\underline{Case 2: $x_{i}+y_{j}< x_{j}+y_{i}$.} For this case, we have
\begin{equation*}
\Delta_{1}(x_{i},x_{j})=u(\min\{x_{i}+y_{j},\min\{(\bm{x}+\bm{y})_{\{i,j\}}\}\})-u(\min\{x_{i}+y_{i},\min\{(\bm{x}+\bm{y})_{\{i,j\}}\}\}).
\end{equation*}
Note that $x_{i}+y_{j}\geq x_{i}+y_{i}$ for any $y_{j}\geq y_{i}$. The proof can be obtained from the following three special cases.

\underline{Subcase 1: $\min\{(\bm{x}+\bm{y})_{\{i,j\}}\}\}\leq x_{i}+y_{i}$.} Clearly, it holds that $\Delta_{1}(x_{i},x_{j})=0$, and the proof is trivial.

\underline{Subcase 2: $x_{i}+y_{i}<\min\{(\bm{x}+\bm{y})_{\{i,j\}}\}\}\leq x_{i}+y_{j}$.} For this case, it is clear to see that
\begin{equation*}
\Delta_{1}(x_{i},x_{j})=u(\min\{(\bm{x}+\bm{y})_{\{i,j\}}\})-u(x_{i}+y_{i})
\end{equation*}
is decreasing in $x_{i}\leq x_{j}$.

\underline{Subcase 3: $\min\{(\bm{x}+\bm{y})_{\{i,j\}}\}\}> x_{i}+y_{j}$.} Note that
\begin{equation*}
\Delta_{1}(x_{i},x_{j})=u(x_{i}+y_{j})-u(x_{i}+y_{i}).
\end{equation*}
By the increasing concavity of $u$, it holds that, for any $x'_{i}\leq x_{i}\leq x_{j}$,
\begin{equation*}
\Delta_{1}(x'_{i},x_{j})=u(x'_{i}+y_{j})-u(x'_{i}+y_{i})\geq u(x_{i}+y_{j})-u(x_{i}+y_{i})=\Delta_{1}(x_{i},x_{j}),
\end{equation*}
which means that $\Delta_{1}(x_{i},x_{j})$ is decreasing in $x_{i}\leq x_{j}$ for any $x_{j}$.

To sum up, we have shown that $g_{2}(x_{i},x_{j})-g_{1}(x_{i},x_{j})$ is decreasing in $x_{i}\leq x_{j}$ for any increasing concave $u$. Now, upon applying Lemma \ref{bivachar}, it follows that
\begin{equation}\label{condieq1}
\mathbb{E}[u(\min(\bm{X}+\bm{y}))|\bm{X}_{\{i,j\}}=\bm{x}_{\{i,j\}}]\leq\mathbb{E}[u(\min(\bm{X}+\tau_{i,j}(\bm{y})))|\bm{X}_{\{i,j\}}=\bm{x}_{\{i,j\}}].
\end{equation}
By applying iterated expectation formula on inequality (\ref{condieq1}), (\ref{condieq2}) is obtained. Upon combining (\ref{condieq2}) with (\ref{densieq1}), (\ref{rwequa1}) is non-negative for $y_{i}\leq y_{j}$. Thus, the proof is finished.

\underline{Proof of (ii): $\bm{X}$ is RWSAI.} According to Proposition 3.9(ii) of Cai and Wei \cite{caiwei14}, the RWSAI property of $\bm{X}$ implies that $[(X_{i},X_{j})|\bm{X}_{\{i,j\}}]$ is RWSAI. For any $y_{i}\leq y_{j}$ and $x_{i}\leq x_{j}$,  $\Delta_{1}(x_{i},x_{j})$ given in (\ref{deltaeq1}) is increasing in $x_{j}\geq x_{i}$ by using the increasing property of $u$. By adopting a similar proof method as in (i), the desired result can be reached.\qed

\medskip

For a series system with components $C_{1},\ldots,C_{n}$ having LWSAI [RWSAI] lifetimes, Theorem \ref{heterseries} suggests that the redundancy $R_{i}$ should be put in standby with $C_{n-i+1}$, for $i=1,\ldots,n$, in the sense of the increasing concave [usual stochastic] ordering if the redundancies lifetimes $Y_i$'s are such that $Y_{1}\geq_{\rm lr}\cdots\geq_{\rm lr}Y_{n}$.

The following example adopts Monte Carlo method to validate the result of Theorem \ref{heterseries}.
\begin{Exa}\label{examin}
\begin{figure}[htbp!]
  \centering
  \subfigure[$u(x)=x^{0.8}$]{
    \label{figmina} 
    \includegraphics[width=0.48\textwidth,height=0.25\textheight]{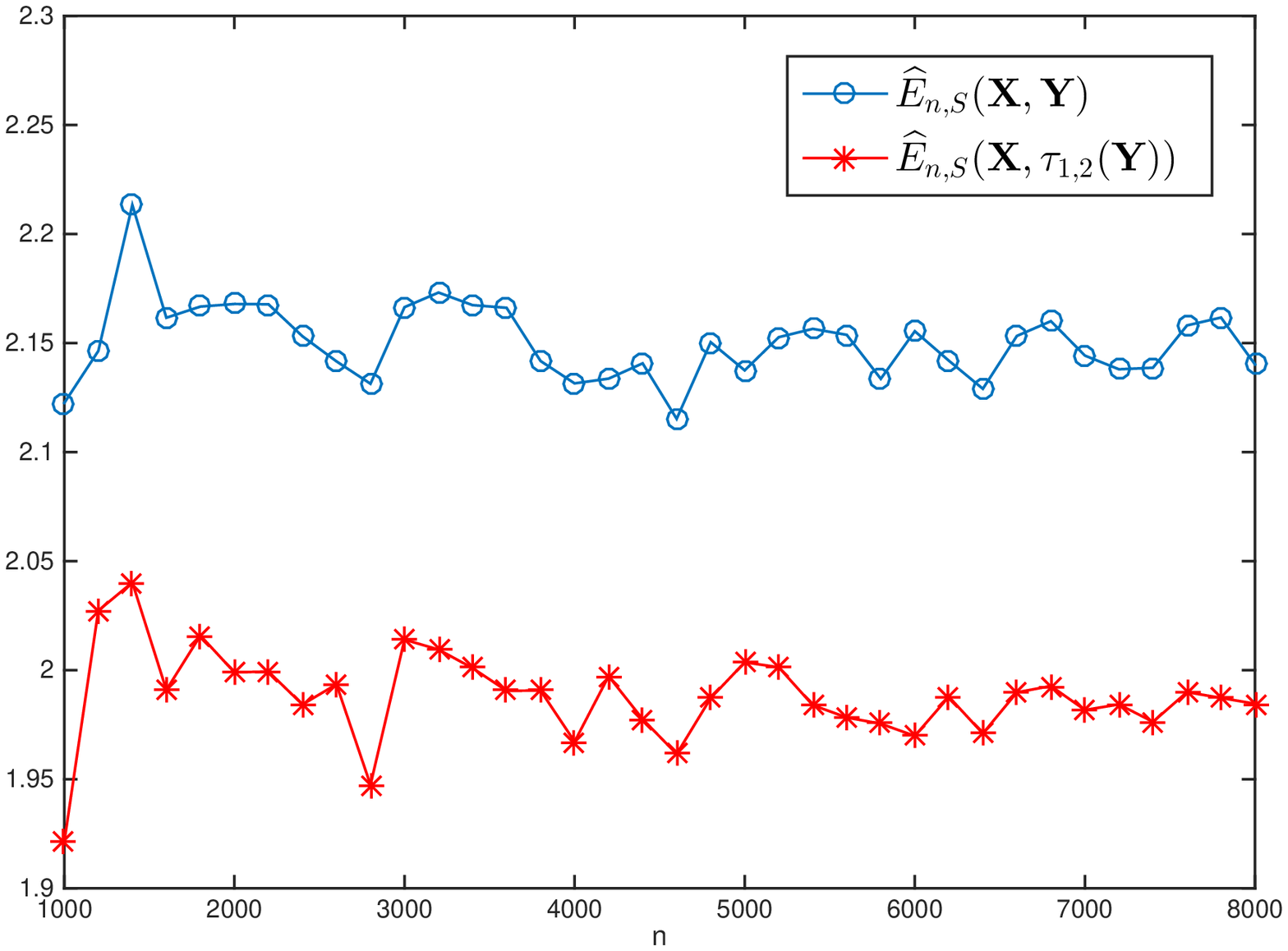}}
  \subfigure[$u(x)=x^{1.2}$]{
    \label{figminb} 
    \includegraphics[width=0.48\textwidth,height=0.25\textheight]{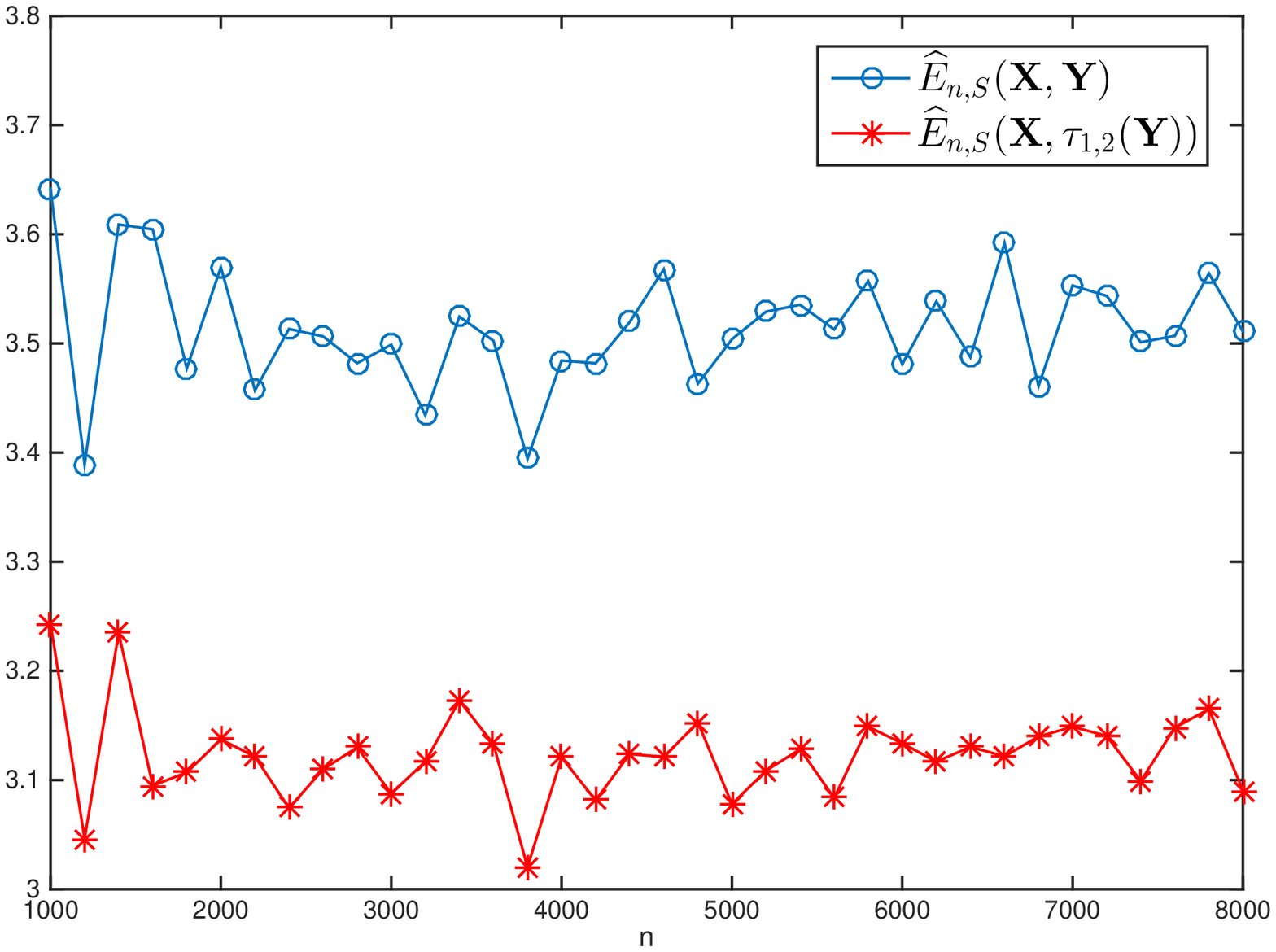}}
    \subfigure[$u(x)=(1-e^{-2x})/2$]{
    \label{figminc} 
    \includegraphics[width=0.48\textwidth,height=0.25\textheight]{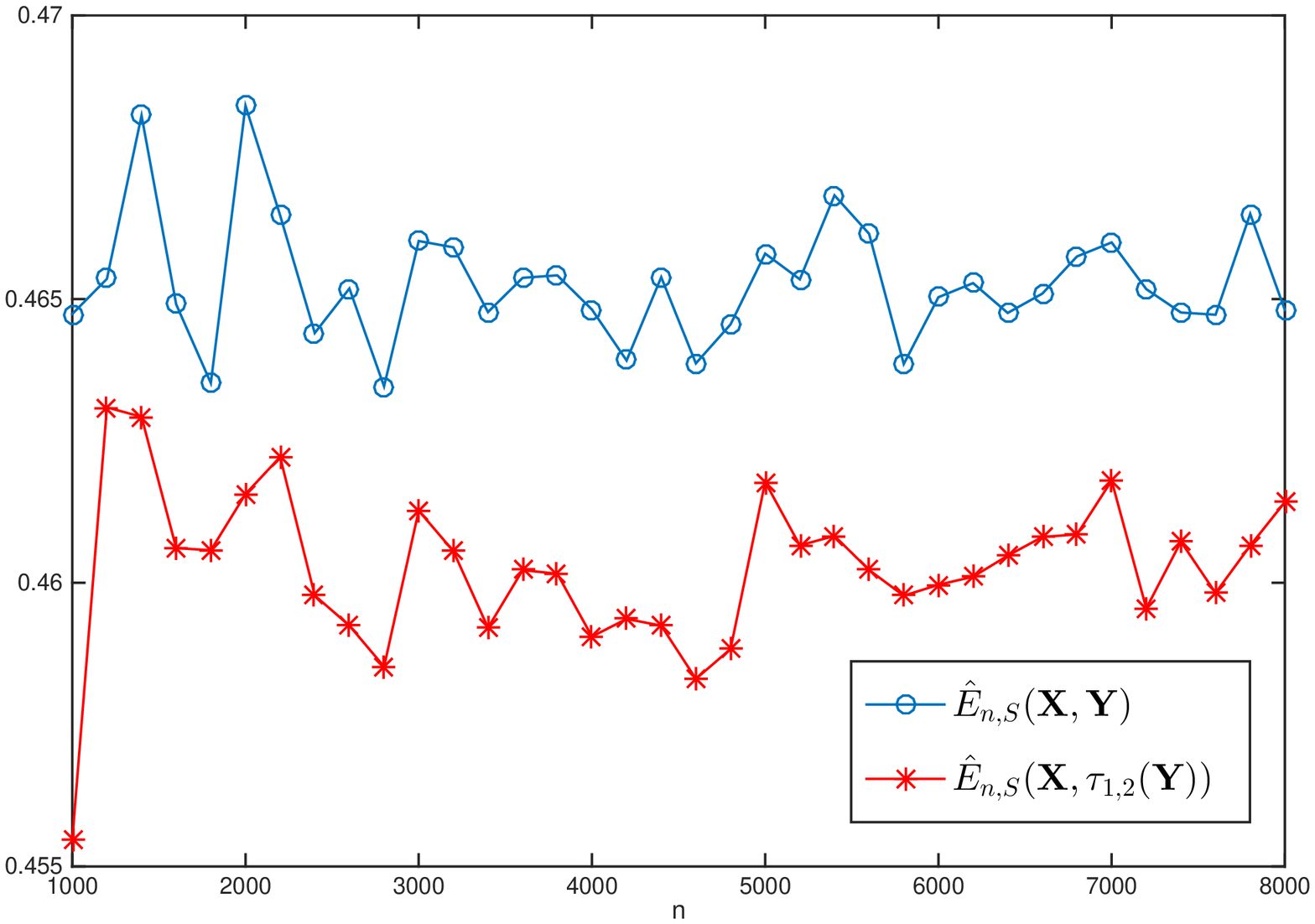}}
    \subfigure[$u(x)=\log x$]{
    \label{figmind} 
    \includegraphics[width=0.48\textwidth,height=0.25\textheight]{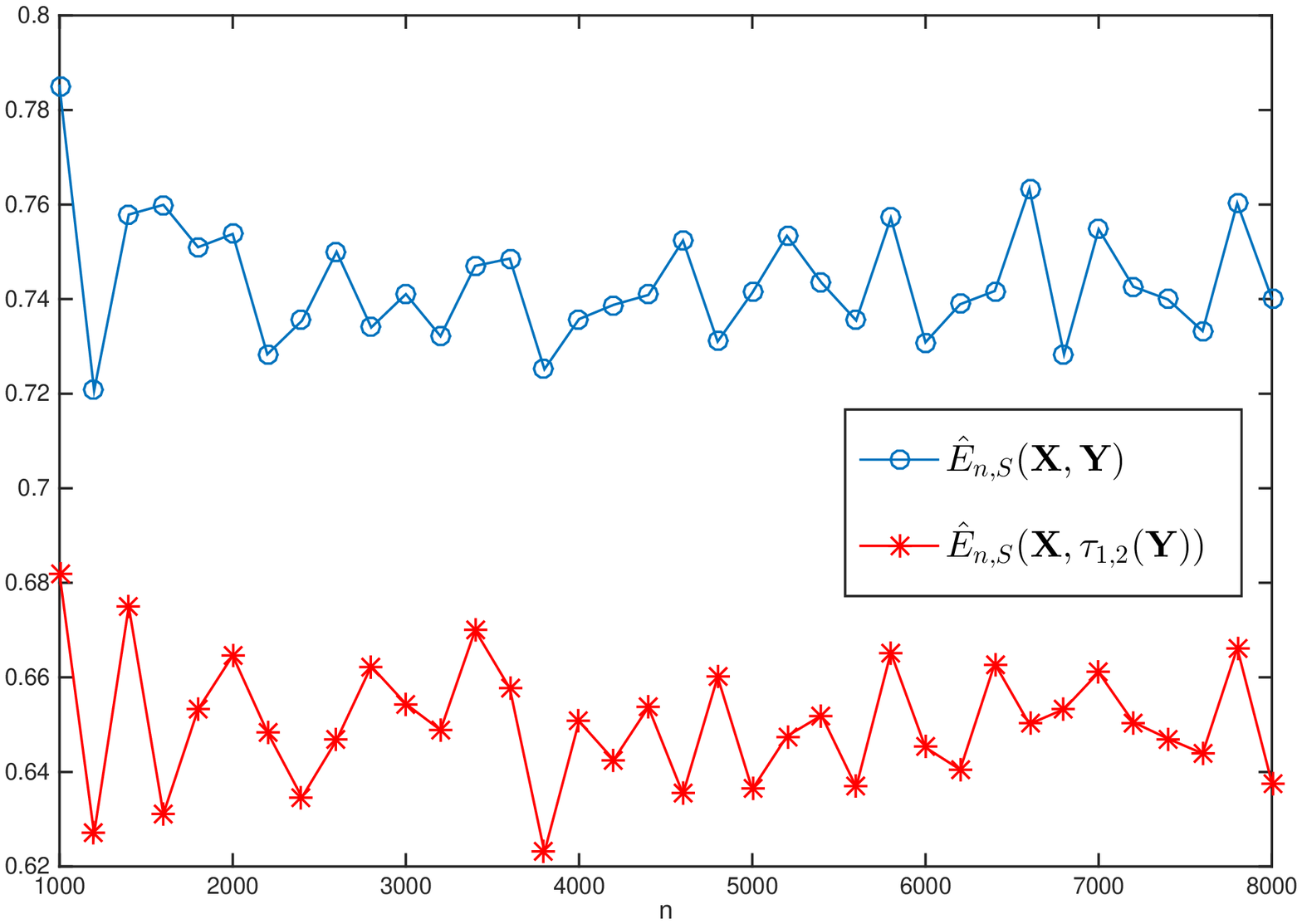}}
  \caption{Plots of the estimators $\widehat{E}_{n,S}(\bm{X},\bm{Y})$ and $\widehat{E}_{n,S}(\bm{X},\tau_{1,2}(\bm{Y}))$.}
  \label{figmin} 
\end{figure}
Assume the lifetime vector $\bm{X}=(X_{1},X_{2})$ is assembled with Clayton [survival] copula with generator $\phi(t)=(t+1)^{-1}$, and $X_{1}$ and $X_{2}$ have exponential distributions with hazard rates $\lambda_{1}=0.8$ and $\lambda_{2}=0.3$, respectively. Let $Y_{1}$ and $Y_{2}$ be two independent exponential random variables with respective hazard rates $\mu_{1}=0.4$ and $\mu_{2}=0.6$. Clearly, $X_{1}\leq_{\rm lr}X_{2}$ and $Y_{1}\geq_{\rm lr}Y_{2}$. Besides, it is easy to check that $\bm{X}$ is LWSAI [RWSAI]. For any increasing function $u$, we denote
\begin{equation*}
E_{u,S}(\bm{X},\bm{Y}):=\mathbb{E}[u(\min\{X_{1}+Y_{1},X_{2}+Y_{2}\})]
\end{equation*}
and
\begin{equation*}
E_{u,S}(\bm{X},\tau_{1,2}(\bm{Y})):=\mathbb{E}[u(\min\{X_{1}+Y_{2},X_{2}+Y_{1}\})].
\end{equation*}
From the population $(X_{1},X_{2},Y_{1},Y_{2})$, we generate i.i.d. samples
\begin{equation*}
(X_{1,1},X_{2,1},Y_{1,1},Y_{2,1}),\ldots,(X_{1,n},X_{2,n},Y_{1,n},Y_{2,n}),
\end{equation*}
{where the generation of $(X_{1,i},X_{2,i})$, for $i=1,\ldots,n$, are based on the method in Subsection 2.9 of Nelsen \cite{nelsen06}.} Then, $E_{u,S}(\bm{X},\bm{Y})$ and $E_{u,S}(\bm{X},\tau_{1,2}(\bm{Y}))$ can be approximated by
\begin{equation*}
\widehat{E}_{n,S}(\bm{X},\bm{Y})=\frac{1}{n}\sum_{i=1}^{n}u(\min\{X_{1,i}+Y_{1,i},X_{2,i}+Y_{2,i}\})
\end{equation*}
and
\begin{equation*}
\widehat{E}_{n,S}(\bm{X},\tau_{1,2}(\bm{Y}))=\frac{1}{n}\sum_{i=1}^{n}u(\min\{X_{1,i}+Y_{2,i},X_{2,i}+Y_{1,i}\}),
\end{equation*}
respectively. {Consider four different utility functions $u(x)=x^{0.8}$, $u(x)=x^{1.2}$, $u(x)=(1-e^{-2x})/2$, and $u(x)=\log x$, for $x\in\mathbb{R}_+$.}  As observed in Figure \ref{figmin}, $\widehat{E}_{n,S}(\bm{X},\bm{Y})\geq\widehat{E}_{n,S}(\bm{X},\tau_{1,2}(\bm{Y}))$ for $n=1000,1200,\ldots,8000$ and all four utility functions. By law of large numbers, we have $E_{u,S}(\bm{X},\bm{Y}) \geq E_{u,S}(\bm{X},\tau_{1,2}(\bm{Y}))$, and thus the result of Theorem \ref{heterseries} is illustrated.
\end{Exa}

As the next example shows, the LWSAI [RWSAI] condition in Theorem \ref{heterseries} may not be discarded.
\begin{Exa}\label{exa01}
\begin{figure}[htbp!]
  \centering
  \subfigure[Distribution functions of $X_1$ and $X_2$]{
    \label{mar} 
    \includegraphics[width=0.48\textwidth,height=0.25\textheight]{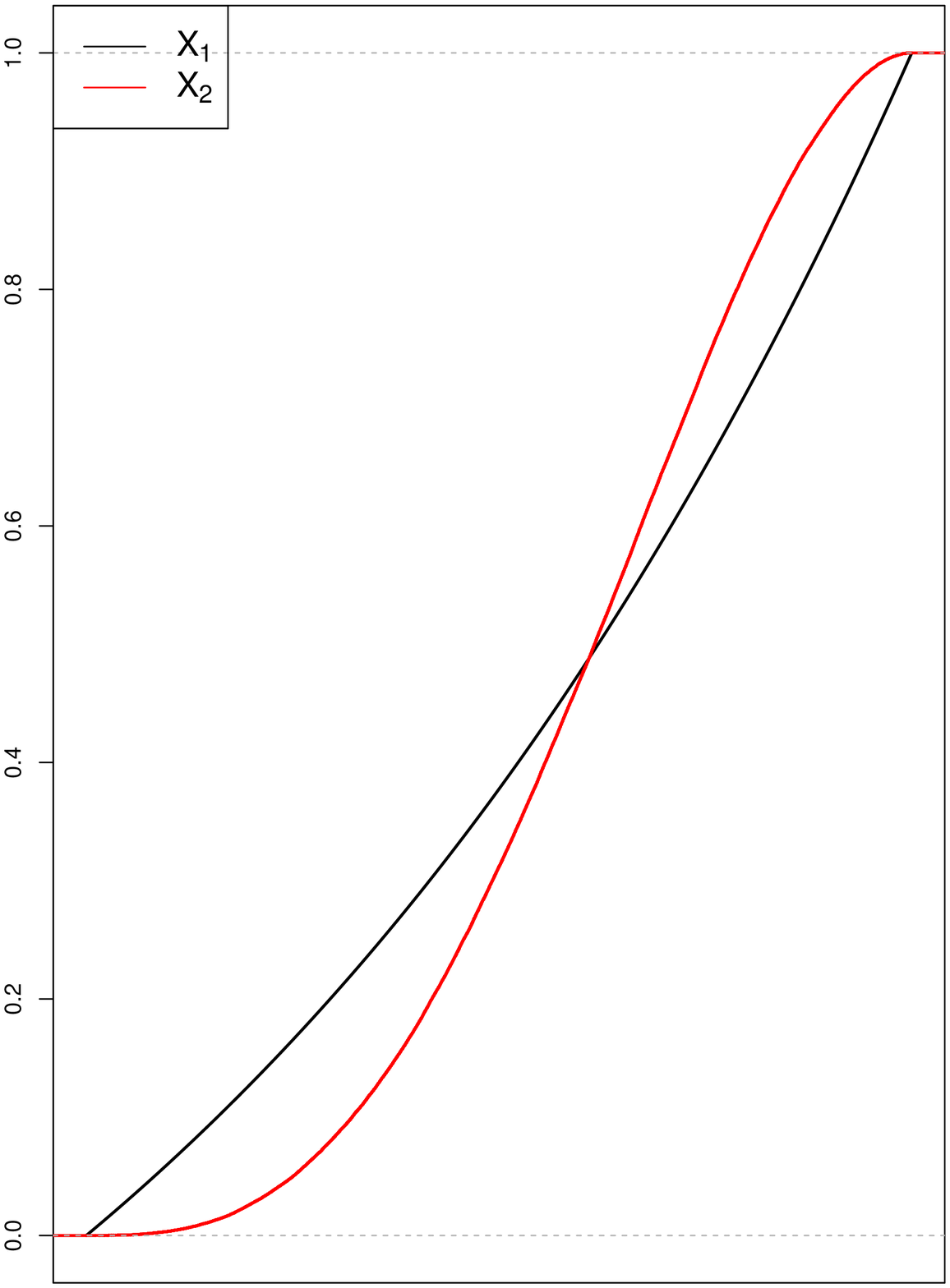}}
  \subfigure[$\widehat{E}_{u,S}(\bm{X},\bm{Y})-\widehat{E}_{u,S}(\bm{X},\tau_{1,2}(\bm{Y}))$]{
    \label{min02} 
    \includegraphics[width=0.48\textwidth,height=0.25\textheight]{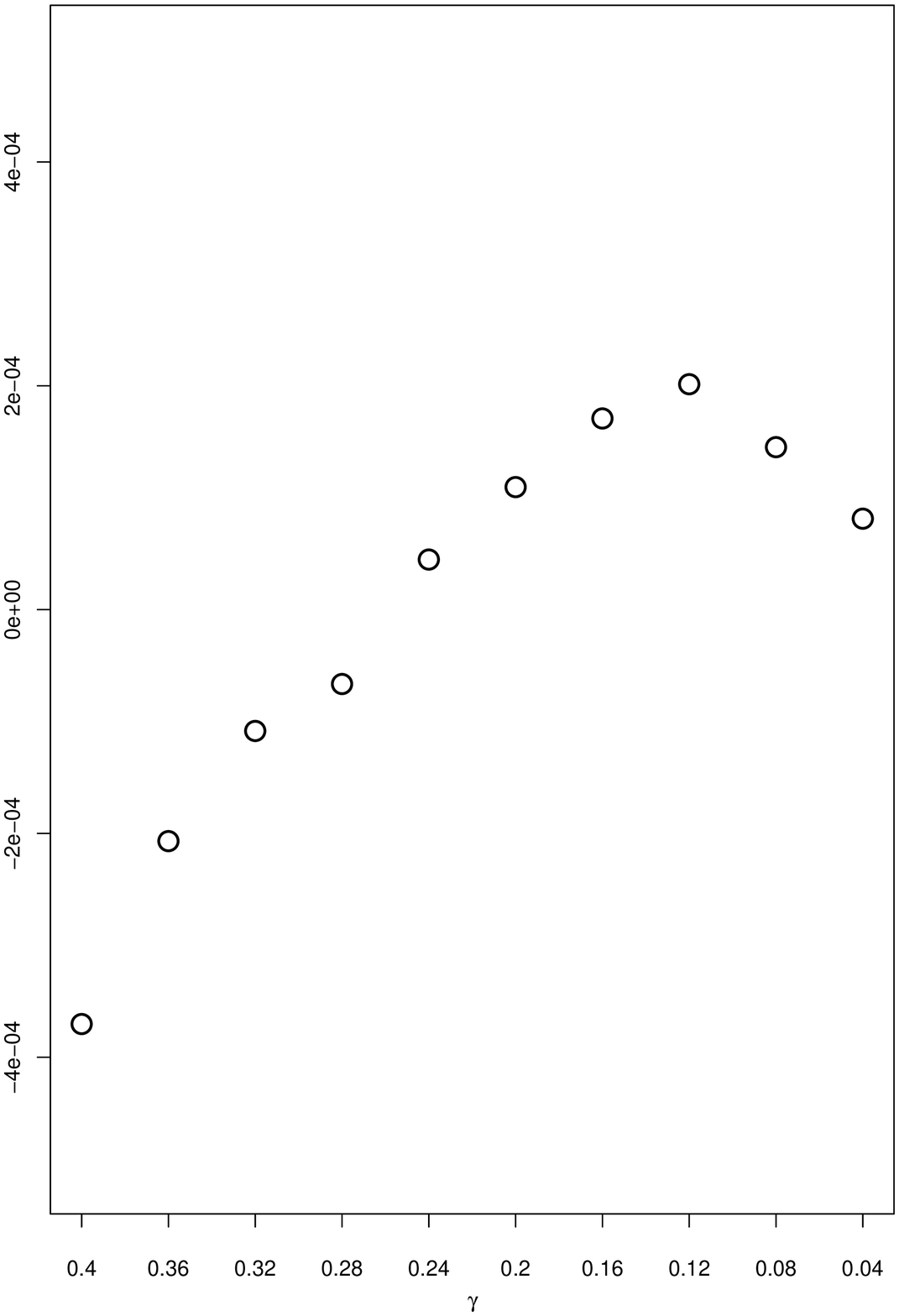}}
  \caption{Plots of Example \ref{exa01}.}
\end{figure}
Assume the lifetime vector $\bm{X}=(X_{1},X_{2}, X_{3})$ is assembled with Gumbel-Barnett copula with generator $\phi(t)=e^{-t^{1/2}}$, and $X_{1}$ has distribution function $(e^x-1)/(e-1)$ for $x\in[0,1]$, $X_{2}$ has beta distribution with two shape parameters $(3,2)$, and $X_3$ has distribution function $(e^{2x}-1)/(e^2-1)$ for $x\in[0,1]$. Let $Y_{1},Y_{2},Y_{3}$ be three independent beta random variables with shape parameters $(3,2)$, $(2,2)$ and $(1,2)$, respectively. One can easily check that $Y_{1}\geq_{\rm lr}Y_{2}\geq_{\rm lr}Y_{3}$. As can be seen in Figure \ref{mar}, the distribution function of $X_1$ and $X_2$ crosses each other, invalidating the existence of the usual stochastic order between these two random variables. By the implication in \eqref{saiimplication}, we know that $\bm{X}$ is neither LWSAI nor RWSAI. For increasing concave functions $u(x)=x^{\gamma}$, we denote
\begin{equation*}
E_{u,S}(\bm{X},\bm{Y}):=\mathbb{E}[(\min\{X_{1}+Y_{1},X_{2}+Y_{2},X_{3}+Y_{3}\})^\gamma]
\end{equation*}
and
\begin{equation*}
E_{u,S}(\bm{X},\tau_{1,2}(\bm{Y})):=\mathbb{E}[(\min\{X_{1}+Y_{2},X_{2}+Y_{1},X_{3}+Y_{3}\})^\gamma].
\end{equation*}
From the population $(X_{1},X_{2},X_{3},Y_{1},Y_{2},Y_{3})$, we generate i.i.d. samples of one million observations
\begin{equation*}
(X_{1,1},X_{2,1},X_{3,1},Y_{1,1},Y_{2,1},Y_{3,1}),\ldots,(X_{1,10^6},X_{2,10^6},X_{3,10^6},Y_{1,10^6},Y_{2,10^6},Y_{3,10^6}).
\end{equation*}
Then, for each $\gamma=0.4,0.36,\cdots,0.04$, $E_{u,S}(\bm{X},\bm{Y})$ and $E_{u,S}(\bm{X},\tau_{1,2}(\bm{Y}))$ can be approximated by
\begin{equation*}
\widehat{E}_{u,S}(\bm{X},\bm{Y})=\frac{1}{10^6}\sum_{i=1}^{10^6}(\min\{X_{1,i}+Y_{1,i},X_{2,i}+Y_{2,i},X_{3,i}+Y_{3,i}\})^{\gamma}
\end{equation*}
and
\begin{equation*}
\widehat{E}_{u,S}(\bm{X},\tau_{1,2}(\bm{Y}))=\frac{1}{10^6}\sum_{i=1}^{10^6}(\min\{X_{1,i}+Y_{2,i},X_{2,i}+Y_{1,i},X_{3,i}+Y_{3,i}\})^{\gamma},
\end{equation*}
respectively. As observed from Figure \ref{min02}, the difference between $\widehat{E}_{u,S}(\bm{X},\bm{Y})$ and $\widehat{E}_{u,S}(\bm{X},\tau_{1,2}(\bm{Y}))$ is not always positive or negative, negating the conclusion of Theorem \ref{heterseries}.
\end{Exa}


Let $Y$ be the lifetime of one single {cold-standby} redundancy $R$. Denote by
\begin{equation*}
  S^{(r)}(\bm{X};Y)=\min(X_{1},\ldots,X_{r-1},X_{r}+Y,X_{r+1},\ldots,X_{n})
\end{equation*}
the resulting lifetime of the series system with $R$ allocated to $C_{r}$, for $r=1,2,\ldots,n$. The following corollary can be obtained from Theorem \ref{heterseries}, which extends Proposition 3.1(i) of Singh and Misra \cite{singh94}, Theorem 5 of Li and Hu \cite{lihu08}, and Theorem 2.1 of Misra et al. \cite{misra11b} to the case of dependent components.
\begin{Cor}\label{COROseries1}
\begin{itemize}
\item [(i)] If $\bm{X}$ is LWSAI, we have $S^{(1)}(\bm{X};Y)\geq_{\rm icv}S^{(2)}(\bm{X};Y)\geq_{\rm icv}\cdots\geq_{\rm icv}S^{(n)}(\bm{X};Y)$.
\item [(ii)] If $\bm{X}$ is RWSAI, we have  $S^{(1)}(\bm{X};Y)\geq_{\rm st}S^{(2)}(\bm{X};Y)\geq_{\rm st}\cdots\geq_{\rm st}S^{(n)}(\bm{X};Y)$.
\end{itemize}
\end{Cor}

Consider two {cold-standby} redundancies $R_{1}, R_{2}$ have lifetimes $Y_{1}, Y_{2}$ such that $Y_{1}\geq_{\rm lr} Y_{2}$. Let $S^{(1,2)}(\bm{X};Y_{1},Y_{2})$ [$S^{(2,1)}(\bm{X};Y_{1},Y_{2})$] be the lifetime of the series system with $C_{1}$ allocated by $R_{1}$ [$R_{2}$] and $C_{2}$ allocated by $R_{2}$ [$R_{1}$]. The following result states that the allocation policy $S^{(1,2)}(\bm{X};Y_{1},Y_{2})$ is better than $S^{(2,1)}(\bm{X};Y_{1},Y_{2})$, which generalizes Theorem 2.2(i) of Li et al. \cite{liyanhu11} and Theorems 3.1 and 3.2 of Misra et al. \cite{misra11b}.
\begin{Cor}
\begin{itemize}
\item [(i)] If $\bm{X}$ is LWSAI, we have $S^{(1,2)}(\bm{X};Y_{1},Y_{2})\geq_{\rm icv}S^{(2,1)}(\bm{X};Y_{1},Y_{2})$.
\item [(ii)] If $\bm{X}$ is RWSAI, we have $S^{(1,2)}(\bm{X};Y_{1},Y_{2})\geq_{\rm st}S^{(2,1)}(\bm{X};Y_{1},Y_{2})$.
\end{itemize}
\end{Cor}

\subsection{Parallel system}
In this subsection, optimal allocation strategies of matched heterogeneous and independent {cold-standby} redundancies are pinpointed for parallel systems comprised of dependent components.
\begin{The}\label{heterparallel} Suppose that $Y_{1}\geq_{\rm lr}Y_{2}\geq_{\rm lr}\cdots\geq_{\rm lr}Y_{n}$. For any $1\leq i<j\leq n$,
\begin{itemize}
\item [(i)]if $\bm{X}$ is LWSAI, then $\max(\bm{X}+\bm{Y})\leq_{\rm st}\max(\bm{X}+\tau_{i,j}(\bm{Y}))$;
\item [(ii)]if $\bm{X}$ is RWSAI, then $\max(\bm{X}+\bm{Y})\leq_{\rm icx}\max(\bm{X}+\tau_{i,j}(\bm{Y}))$.
\end{itemize}
\end{The}
\proof By adopting the proof of Theorem \ref{heterseries}, it suffices to show the non-positivity of
\begin{eqnarray}\label{pareq1}
\lefteqn{\mathbb{E}[u(\max(\bm{X}+\bm{Y}))]-\mathbb{E}[u(\max(\bm{X}+\tau_{i,j}(\bm{Y})))]}&& \nonumber\\
  &=&\idotsint\limits_{\mathbb{R}_{+}^{n-2}}\prod_{k\neq i,j}^{n}p_{k}(y_{k}) \dif y_{k}\iint\limits_{y_{i}\leq y_{j}}[\mathbb{E}[u(\max(\bm{X}+\bm{y}))]-\mathbb{E}[u(\max(\bm{X}+\tau_{i,j}(\bm{y})))]]\nonumber\\
  &&\quad\times[p_{i}(y_{i})p_{j}(y_{j})-p_{i}(y_{j})p_{j}(y_{i})]\dif y_{i}\dif y_{j},
\end{eqnarray}
where $u$ is increasing for case (i), and increasing convex for case (ii). In view of (\ref{densieq1}), it is enough to prove that
\begin{equation}\label{maxeq2}
\mathbb{E}[u(\max(\bm{X}+\bm{y}))]\geq\mathbb{E}[u(\max(\bm{X}+\tau_{i,j}(\bm{y})))],\quad y_{i}\leq y_{j}.
\end{equation}

\underline{Proof of (i): $\bm{X}$ is LWSAI.} For any given $\bm{X}_{\{i,j\}}=\bm{x}_{\{i,j\}}$, we define
\begin{equation*}
g_{2}(x_{i},x_{j})=u(\max(\bm{x}+\bm{y}))=u(\max\{x_{i}+y_{i},x_{j}+y_{j},\max\{(\bm{x}+\bm{y})_{\{i,j\}}\}\})
\end{equation*}
and
\begin{equation*}
g_{1}(x_{i},x_{j})=u(\max(\bm{x}+\tau_{i,j}(\bm{y})))=u(\max\{x_{i}+y_{j},x_{j}+y_{i},\max\{(\bm{x}+\bm{y})_{\{i,j\}}\}\}).
\end{equation*}
Then, for any $x_{i}\leq x_{j}$ and $y_{i}\leq y_{j}$, it can be seen that
\begin{eqnarray*}
\Delta_{2}(x_{i},x_{j})&=&g_{2}(x_{i},x_{j})-g_{1}(x_{i},x_{j})\\
&=& u(\max\{x_{j}+y_{j},\max\{(\bm{x}+\bm{y})_{\{i,j\}}\}\})\\
&&\quad -u(\max\{x_{i}+y_{j},x_{j}+y_{i},\max\{(\bm{x}+\bm{y})_{\{i,j\}}\}\})
\end{eqnarray*}
is always decreasing in $x_{i}\leq x_{j}$ for any increasing $u$. On the other hand, for any $x_{i}\leq x_{j}$,
\begin{equation}\label{ln2}
g_{2}(x_{i},x_{j})+g_{2}(x_{j},x_{i})=g_{1}(x_{i},x_{j})+g_{1}(x_{j},x_{i}).
\end{equation}
Thus, we have
\begin{equation*}
\mathbb{E}[u(\max(\bm{X}+\bm{y}))|\bm{X}_{\{i,j\}}=\bm{x}_{\{i,j\}}]\geq\mathbb{E}[u(\max(\bm{X}+\tau_{i,j}(\bm{y})))|\bm{X}_{\{i,j\}}=\bm{x}_{\{i,j\}}].
\end{equation*}
Then, the proof is completed by applying Lemma \ref{bivachar} and iterated expectation formula.

\underline{Proof of (ii): $\bm{X}$ is RWSAI.} In light of the proof in (i), the desired result boils down to showing that $\Delta_{2}(x_{i},x_{j})$ is increasing in $x_{j}\geq x_{i}$ for any $y_{i}\leq y_{j}$ and increasing convex $u$.

\underline{Case 1: $x_{i}+y_{j}\geq x_{j}+y_{i}$.} For this case, the function
\begin{equation*}
\Delta_{2}(x_{i},x_{j})=u(\max\{x_{j}+y_{j},\max\{(\bm{x}+\bm{y})_{\{i,j\}}\}\})-u(\max\{x_{i}+y_{j},\max\{(\bm{x}+\bm{y})_{\{i,j\}}\}\})
\end{equation*}
is always increasing in $x_{j}\geq x_{i}$ by the increasing property of $u$.

\underline{Case 2: $x_{i}+y_{j}< x_{j}+y_{i}$.} For this case, we have
\begin{equation*}
\Delta_{2}(x_{i},x_{j})=u(\max\{x_{j}+y_{j},\max\{(\bm{x}+\bm{y})_{\{i,j\}}\}\})-u(\max\{x_{j}+y_{i},\max\{(\bm{x}+\bm{y})_{\{i,j\}}\}\}).
\end{equation*}
Note that $x_{j}+y_{j}\geq x_{j}+y_{i}$, the following three situations are considered.

\underline{Subcase 1: $\max\{(\bm{x}+\bm{y})_{\{i,j\}}\}\geq x_{j}+y_{j}$.} Clearly, $\Delta_{2}(x_{i},x_{j})=0$, due to which the proof is trivial.

\underline{Subcase 2: $x_{j}+y_{j}>\max\{(\bm{x}+\bm{y})_{\{i,j\}}\}\geq x_{j}+y_{i}$.} Note that
\begin{equation*}
\Delta_{2}(x_{i},x_{j})=u(x_{j}+y_{j})-u(\max\{(\bm{x}+\bm{y})_{\{i,j\}}\}),
\end{equation*}
which is obviously increasing in $x_{j}\geq x_{i}$ by the increasing property of $u$.

\underline{Subcase 3: $x_{j}+y_{i}>\max\{(\bm{x}+\bm{y})_{\{i,j\}}\}$.} Observe that
\begin{equation*}
\Delta_{2}(x_{i},x_{j})=u(x_{j}+y_{j})-u(x_{j}+y_{i}).
\end{equation*}
For any $x_{j}\geq x'_{j}\geq x_{i}$, the increasing convexity of $u$ implies that
\begin{equation*}
\Delta_{2}(x_{i},x_{j})=u(x_{j}+y_{j})-u(x_{j}+y_{i})\geq u(x'_{j}+y_{j})-u(x'_{j}+y_{i})=\Delta_{2}(x_{i},x'_{j}),
\end{equation*}
which means that $\Delta_{2}(x_{i},x_{j})$ is increasing in $x_{j}\geq x_{i}$.

To conclude, we have shown that $\Delta_{2}(x_{i},x_{j})$ is increasing in $x_{j}\geq x_{i}$ for any $y_{i}\leq y_{j}$ and increasing convex $u$. Now, upon using Lemma \ref{bivachar} and iterated expectation formula, the proof is finished.\qed

\medskip

For the parallel system with components $C_{1},\ldots,C_{n}$ having LWSAI [RWSAI] lifetimes, Theorem \ref{heterparallel} implies that the redundancy $R_{i}$ should be put in standby with $C_{i}$, $i=1,\ldots,n$, according to the usual stochastic [increasing convex] ordering if the redundancies lifetimes satisfy $Y_{1}\geq_{\rm lr}\cdots\geq_{\rm lr}Y_{n}$.

For one single {cold-standby} redundancy $R$ with lifetime $Y$, let
\begin{equation*}
  T^{(r)}(\bm{X};Y)=\max(X_{1},\ldots,X_{r-1},X_{r}+Y,X_{r+1},\ldots,X_{n})
\end{equation*}
be the resulting lifetime of the parallel system with $R$ allocated to $C_{r}$, for $r=1,2,\ldots,n$. The following corollary can be obtained from Theorem \ref{heterparallel}, extending Proposition 3.1(ii) of Singh and Misra \cite{singh94}, Theorem 4 of Li and Hu \cite{lihu08}, and Theorem 2.2 of Misra et al. \cite{misra11b} to the case of dependent components.
\begin{Cor}\label{COROparallel1}
\begin{itemize}
\item [(i)] If $\bm{X}$ is LWSAI, we have $T^{(1)}(\bm{X};Y)\leq_{\rm st}T^{(2)}(\bm{X};Y)\leq_{\rm st}\cdots\leq_{\rm st}T^{(n)}(\bm{X};Y)$.
\item [(ii)] If $\bm{X}$ is RWSAI, we have $T^{(1)}(\bm{X};Y)\leq_{\rm icx}T^{(2)}(\bm{X};Y)\leq_{\rm icx}\cdots\leq_{\rm icx}T^{(n)}(\bm{X};Y)$.
\end{itemize}
\end{Cor}

Suppose there are two {cold-standby} redundancies $R_{1}, R_{2}$ having lifetimes $Y_{1}, Y_{2}$ such that $Y_{1}\geq_{\rm lr} Y_{2}$. Let $T^{(1,2)}(\bm{X};Y_{1},Y_{2})$ [$T^{(2,1)}(\bm{X};Y_{1},Y_{2})$] be the resulting lifetime of the parallel system with $C_{1}$ allocated by $R_{1}$ [$R_{2}$] and $C_{2}$ allocated by $R_{2}$ [$R_{1}$]. The following result implies that the allocation policy $T^{(2,1)}(\bm{X};Y_{1},Y_{2})$ is better than $T^{(1,2)}(\bm{X};Y_{1},Y_{2})$, which generalizes Lemma 3.2 of Shaked and Shanthikumar \cite{ss07}, Theorem 2.2(ii) of Li et al. \cite{liyanhu11}, and Theorem 3.3 of Misra et al. \cite{misra11b}.
\begin{Cor}
\begin{itemize}
\item [(i)] If $\bm{X}$ is LWSAI, we have $T^{(1,2)}(\bm{X};Y_{1},Y_{2})\leq_{\rm st}T^{(2,1)}(\bm{X};Y_{1},Y_{2})$.
\item [(ii)] If $\bm{X}$ is RWSAI, we have $T^{(1,2)}(\bm{X};Y_{1},Y_{2})\leq_{\rm icx}T^{(2,1)}(\bm{X};Y_{1},Y_{2})$.
\end{itemize}
\end{Cor}

To close this section, we present one example to illustrate Theorem \ref{heterparallel}.
\begin{Exa} Under the setting of Example \ref{examin}, we denote
\begin{equation*}
E_{u,P}(\bm{X},\bm{Y}):=\mathbb{E}[u(\max\{X_{1}+Y_{1},X_{2}+Y_{2}\})]
\end{equation*}
and
\begin{equation*}
E_{u,P}(\bm{X},\tau_{1,2}(\bm{Y})):=\mathbb{E}[u(\max\{X_{1}+Y_{2},X_{2}+Y_{1}\})].
\end{equation*}
Then, the function $E_{u,P}(\bm{X},\bm{Y})$ and $E_{u,P}(\bm{X},\tau_{1,2}(\bm{Y}))$ can be approximated by
\begin{equation*}
\widehat{E}_{n,P}(\bm{X},\bm{Y})=\frac{1}{n}\sum_{i=1}^{n}u(\max\{X_{1,i}+Y_{1,i},X_{2,i}+Y_{2,i}\})
\end{equation*}
and
\begin{equation*}
\widehat{E}_{n,P}(\bm{X},\tau_{1,2}(\bm{Y}))=\frac{1}{n}\sum_{i=1}^{n}u(\max\{X_{1,i}+Y_{2,i},X_{2,i}+Y_{1,i}\}),
\end{equation*}
respectively, where $(X_{1,i},X_{2,i},Y_{1,i},Y_{2,i})$'s are independent copies of $(X_{1},X_{2},Y_{1},Y_{2})$. {We consider four different utility functions $u(x)=x^{0.8}$, $u(x)=x^{1.2}$, $u(x)=10(1-e^{-0.1x})$, and $u(x)=\log x$. Figure \ref{figmax} shows that $\widehat{E}_{n,P}(\bm{X},\bm{Y})\leq\widehat{E}_{n,P}(\bm{X},\tau_{1,2}(\bm{Y}))$ for $n=1000,1200,\ldots,8000$ and all four types of utility functions}. By law of large numbers, it must hold that $E_{u,P}(\bm{X},\bm{Y})\leq E_{u,P}(\bm{X},\tau_{1,2}(\bm{Y}))$, which validates the effectiveness of Theorem \ref{heterparallel}.
\begin{figure}[htp!]
  \centering
  \subfigure[$u(x)=x^{0.8}$]{
    \label{figmaxa} 
    \includegraphics[width=0.48\textwidth,height=0.25\textheight]{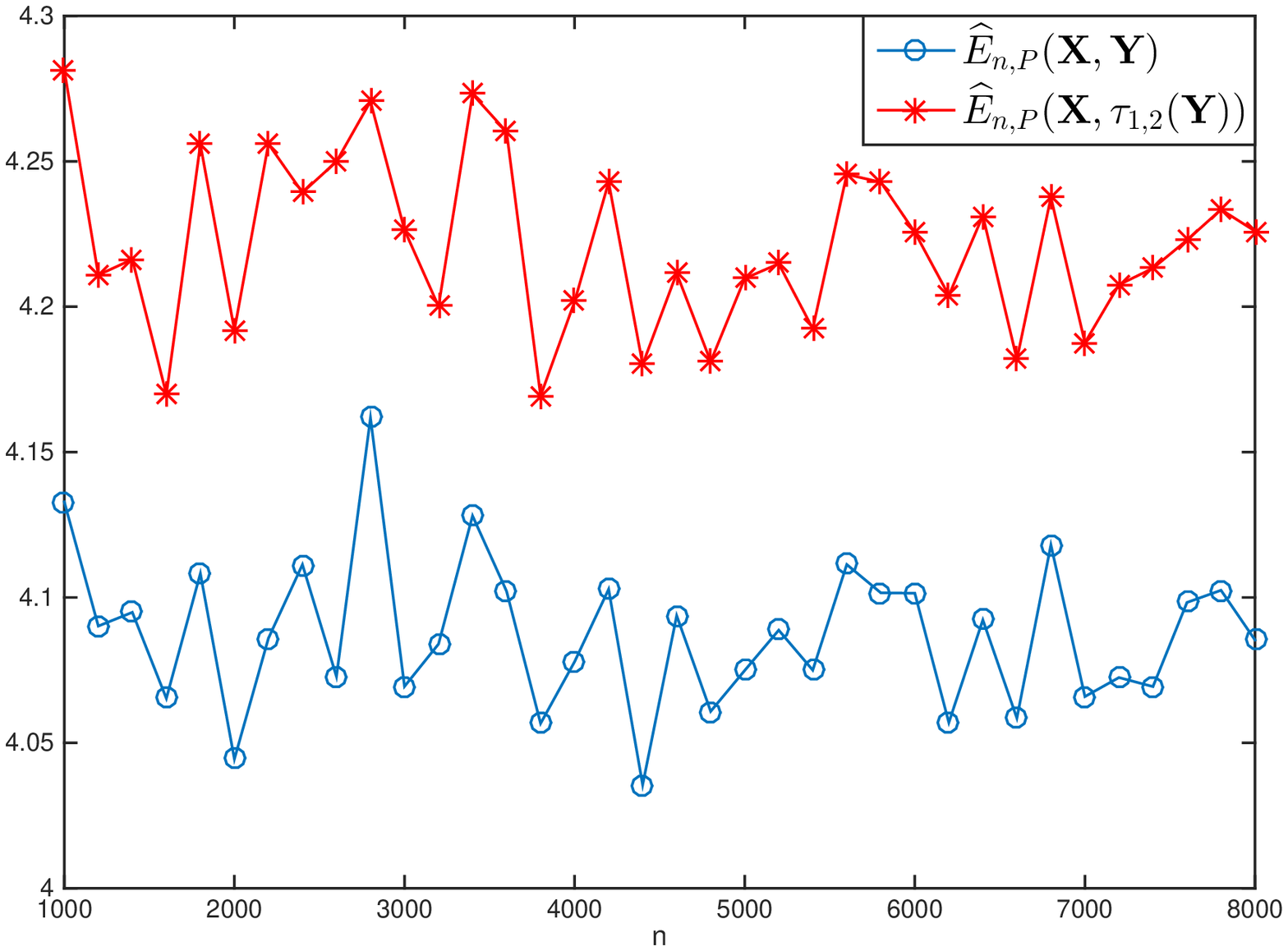}}
  \subfigure[$u(x)=x^{1.2}$]{
    \label{figmaxb} 
    \includegraphics[width=0.48\textwidth,height=0.25\textheight]{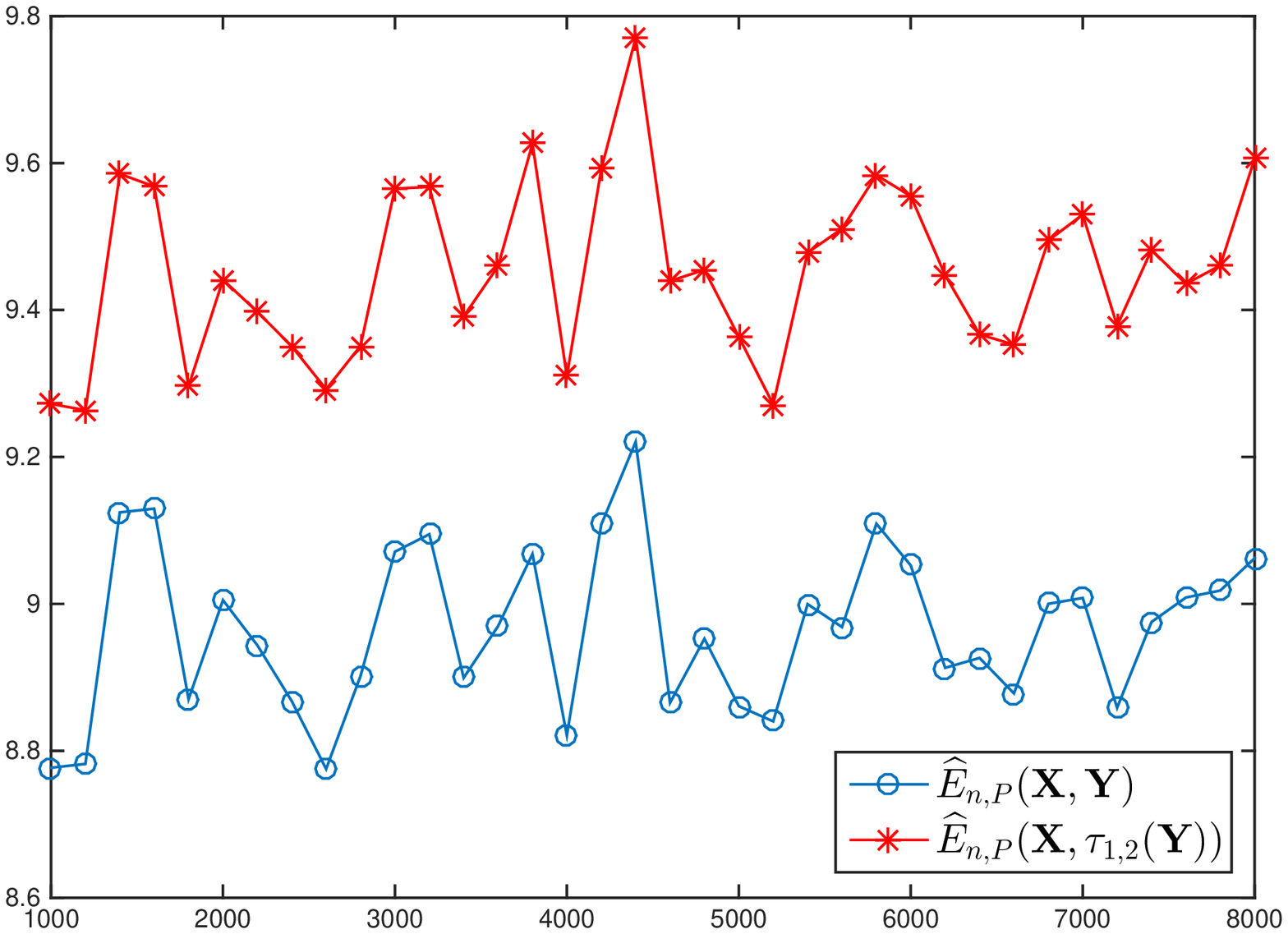}}
    \subfigure[$u(x)=10(1-e^{-0.1x})$]{
    \label{figmaxc} 
    \includegraphics[width=0.48\textwidth,height=0.25\textheight]{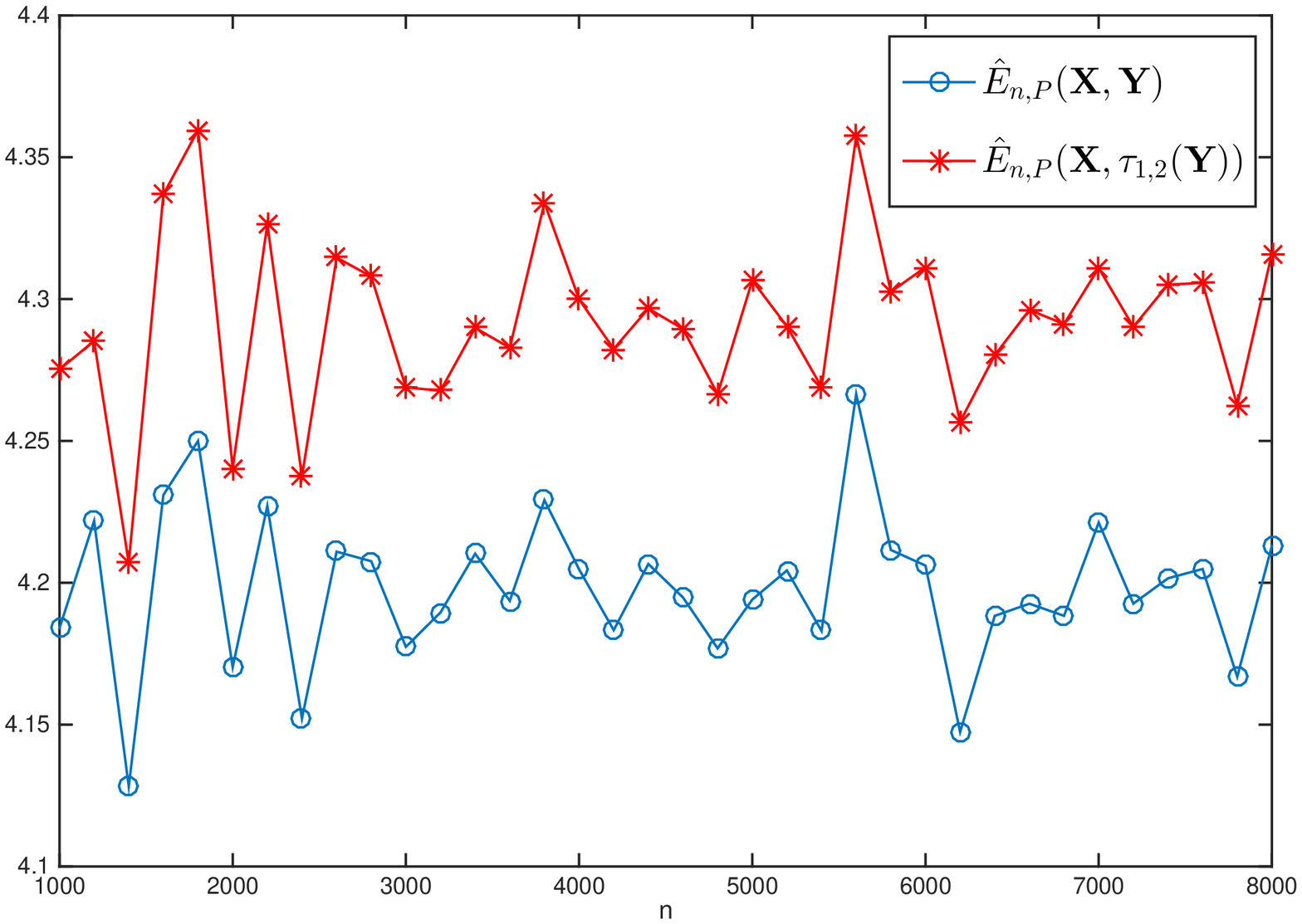}}
     \subfigure[$u(x)=\log x$]{
    \label{figmaxc} 
    \includegraphics[width=0.48\textwidth,height=0.25\textheight]{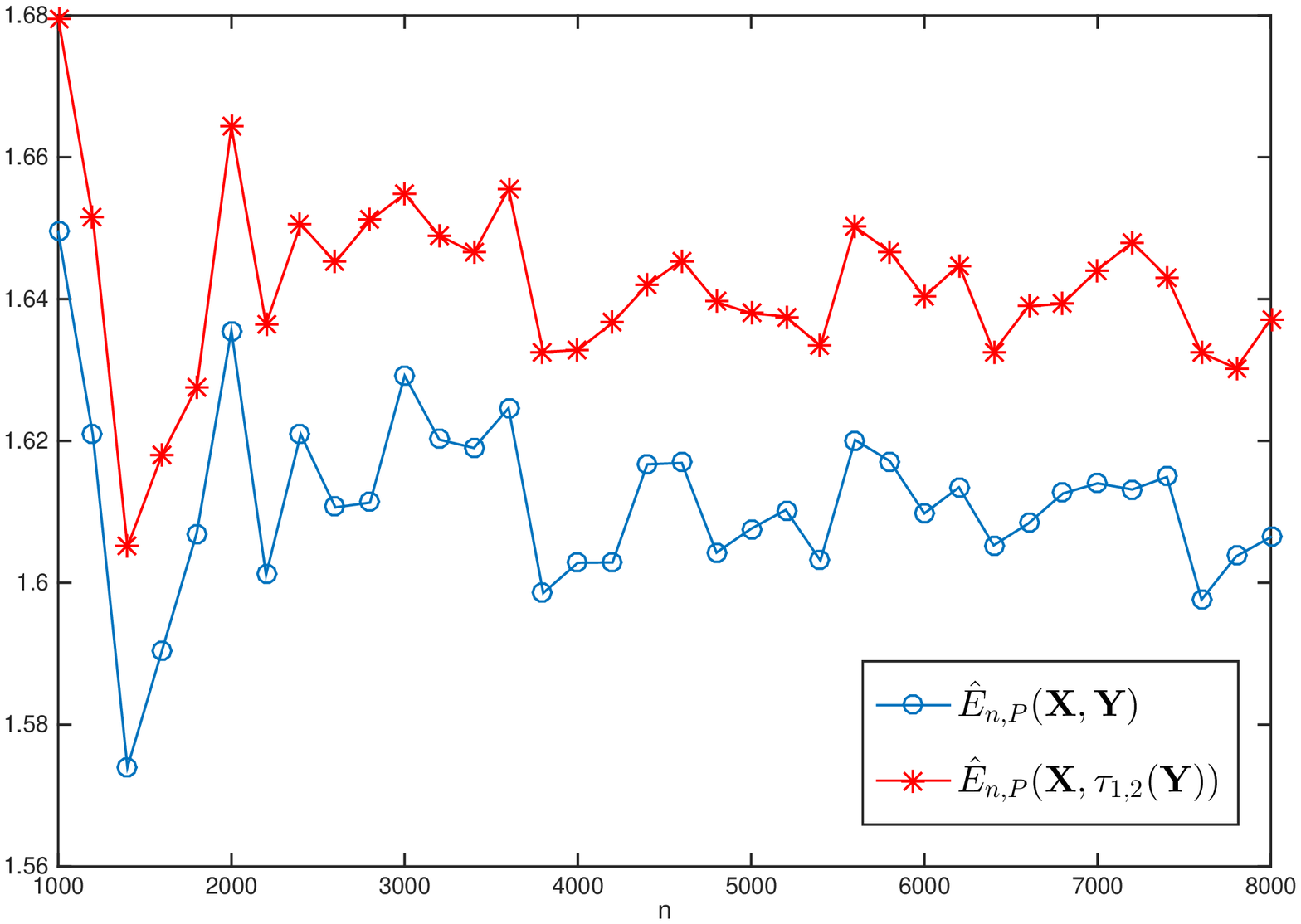}}
  \caption{Plots of the estimators $\widehat{E}_{n,P}(\bm{X},\bm{Y})$ and $\widehat{E}_{n,P}(\bm{X},\tau_{1,2}(\bm{Y}))$.}
  \label{figmax} 
\end{figure}
\end{Exa}

\begin{Exa} Under the setting of Example \ref{exa01}, we denote
\begin{equation*}
E_{u,P}(\bm{X},\bm{Y}):=\mathbb{E}[(\max\{X_{1}+Y_{1},X_{2}+Y_{2},X_{3}+Y_{3}\})^{\gamma}]
\end{equation*}
and
\begin{equation*}
E_{u,P}(\bm{X},\tau_{1,2}(\bm{Y})):=\mathbb{E}[(\max\{X_{1}+Y_{2},X_{2}+Y_{1},X_{3}+Y_{3}\})^{\gamma}].
\end{equation*}
Then, the function $E_{u,P}(\bm{X},\bm{Y})$ and $E_{u,P}(\bm{X},\tau_{1,2}(\bm{Y}))$ can be approximated by
\begin{equation*}
\widehat{E}_{u,P}(\bm{X},\bm{Y})=\frac{1}{10^6}\sum_{i=1}^{10^6}(\max\{X_{1,i}+Y_{1,i},X_{2,i}+Y_{2,i},X_{3,i}+Y_{3,i}\})^{\gamma}
\end{equation*}
and
\begin{equation*}
\widehat{E}_{u,P}(\bm{X},\tau_{1,2}(\bm{Y}))=\frac{1}{10^6}\sum_{i=1}^{10^6}(\max\{X_{1,i}+Y_{2,i},X_{2,i}+Y_{1,i},X_{3,i}+Y_{3,i}\})^{\gamma},
\end{equation*}
respectively. Figure \ref{figmax02} shows the difference between $\widehat{E}_{u,P}(\bm{X},\bm{Y})$ and $\widehat{E}_{u,P}(\bm{X},\tau_{1,2}(\bm{Y}))$ for $\gamma=1,1.2,\cdots,3$. As can be seen from the plot, the difference curve is neither positive nor negative for all the $\gamma$, which invalidates the effectiveness of Theorem \ref{heterparallel}.
\begin{figure}[htp!]
  \centering
    \includegraphics[width=0.48\textwidth,height=0.25\textheight]{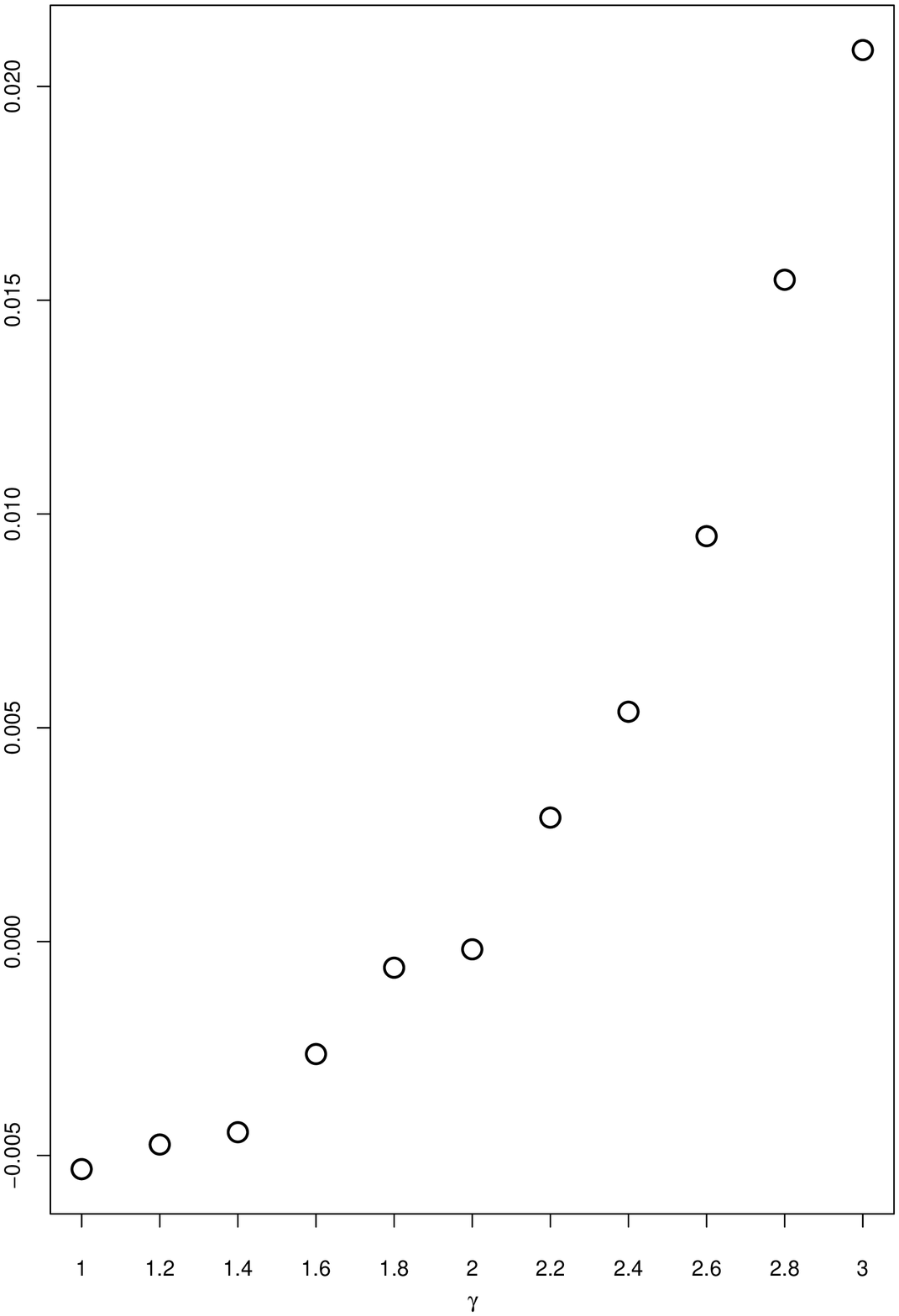}
  \caption{Difference between $\widehat{E}_{u,P}(\bm{X},\bm{Y})$ and $\widehat{E}_{u,P}(\bm{X},\tau_{1,2}(\bm{Y}))$.}
  \label{figmax02} 
\end{figure}
\end{Exa}

{One natural interesting problem would be studying optimal {cold-standby} redundancies allocations for general $k$-out-of-$n$ systems. The following example indicates that there is no certain answer for this problem.
\begin{figure}[htp!]
  \centering
  \subfigure[$\lambda=0.4$]{
    \label{figfail1} 
    \includegraphics[width=0.48\textwidth,height=0.25\textheight]{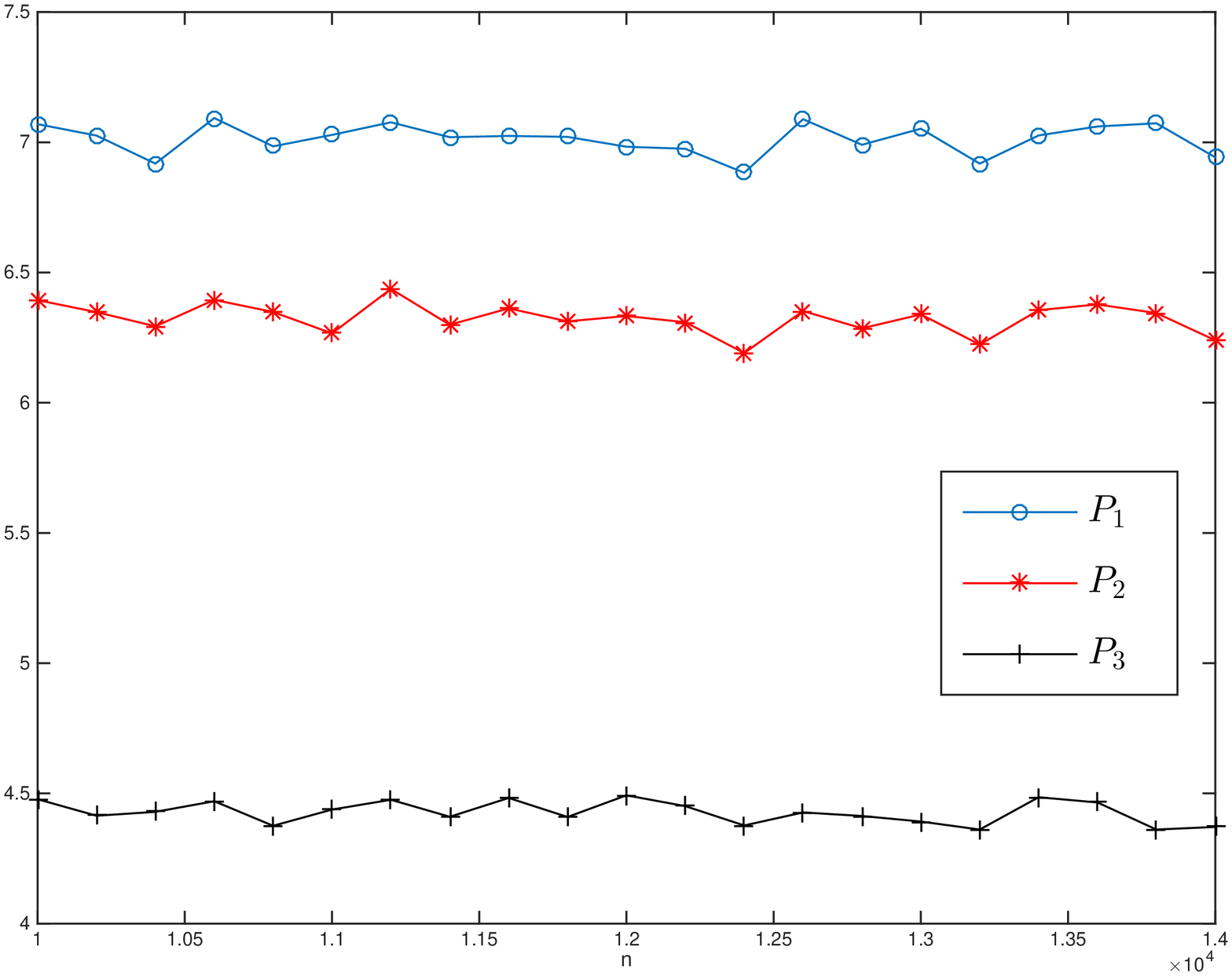}}
  \subfigure[$\lambda=2.2$]{
    \label{figfail2} 
    \includegraphics[width=0.48\textwidth,height=0.25\textheight]{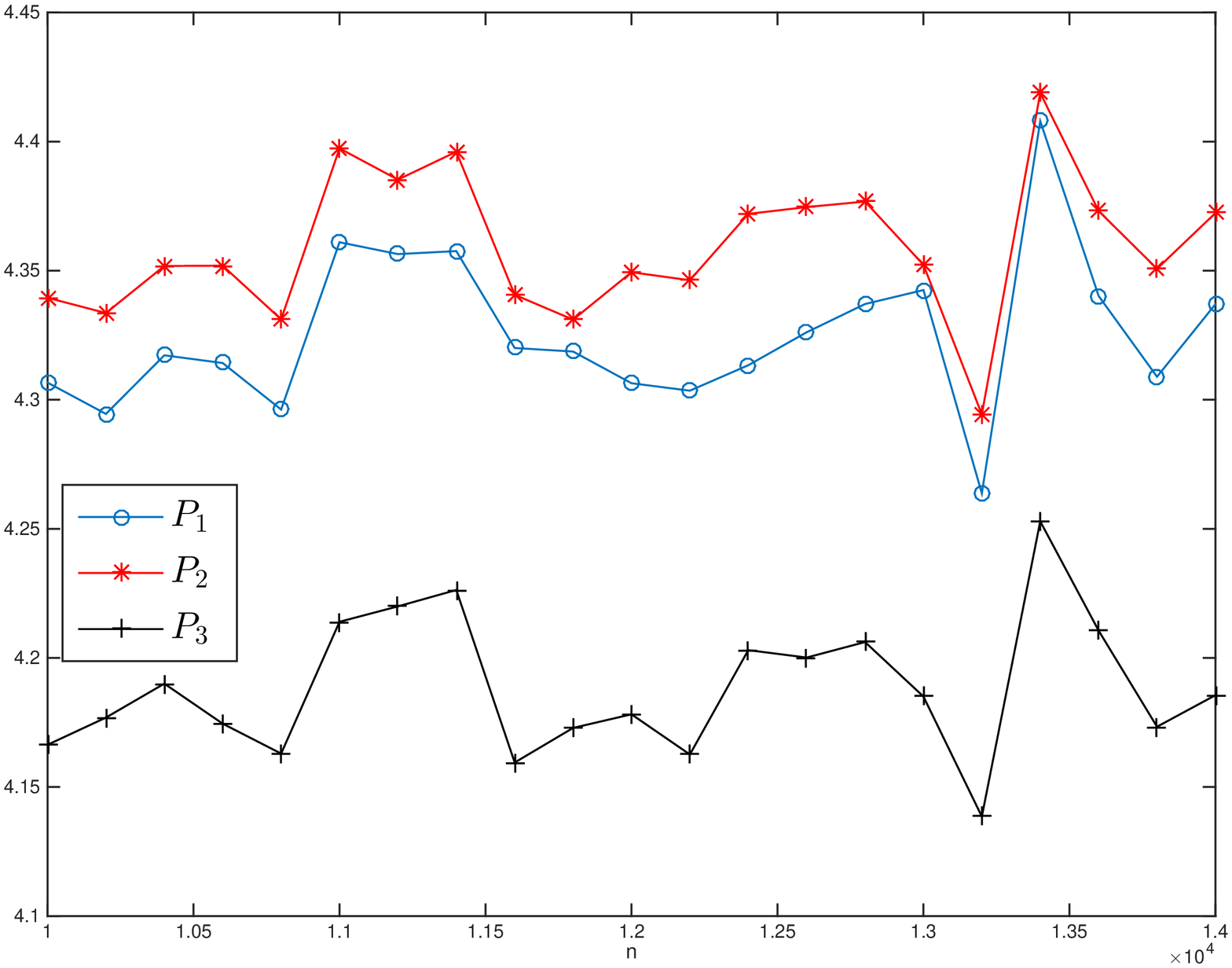}}
  \caption{Plots of $\widehat{E}_{3}(P_i)$ for different allocation policies $P_i$, $i=1,2,3$.}
  \label{figfailsafe} 
\end{figure}
\begin{Exa} Consider a $2$-out-of-$3$ system comprising three independent components with $X_{1}$, $X_{2}$ and $X_{3}$ having hazard rates $\lambda_{1}=0.5$, $\lambda_{2}=0.4$ and $\lambda_{3}=0.3$, respectively. Suppose the {cold-standby} redundancy has exponential lifetime $Y$ with hazard rate $\lambda$. Let us consider three possible allocation policies (i) $P_1$: allocated to $X_{1}$; (ii) $P_2$: allocated to $X_{2}$; and (iii) $P_3$: allocated to $X_{3}$. By taking $u(x)=u^{1.2}$ and $\lambda=0.4$ or $\lambda=2.2$, Figure \ref{figfailsafe} presents the empirical values (denoted by $\widehat{E}_{3}(P_i)$) of the expected function of the resulting lifetime under policies $P_i$, for $i=1,2,3$. As observed from the plots, the optimal allocation policy (in the sense of the increasing convex order) may depend on the reliability performance of the redundancy.
\end{Exa}}

\section{Homogeneous {cold-standby} redundancies}\label{sechomo}
In this section, we investigate optimal allocations of $m$ i.i.d. {cold-standby} redundancies with random lifetimes $\bm{Y}=(Y_{1},\ldots,Y_{m})$ to a series or parallel system comprised of $n$ dependent components with lifetimes $\bm{X}=(X_{1},\ldots,X_{n})$. Let $\bm{r}\in\mathcal{A}:=\{(r_{1},\ldots,r_{n})|\sum_{i=1}^{n}r_{i}=m, r_{i}\in\mathbb{N}, i=1,\ldots,n\}$ be the allocation policy with $r_{i}$ {cold-standby} redundancies allocated to component $C_{i}$, $i=1,\ldots,n$. Denote by $S(\bm{X}+\bm{Y};\bm{r})$ $[T(\bm{X}+\bm{Y};\bm{r})]$ the lifetime of the series [parallel] system with allocation policy $\bm{r}\in\mathcal{A}$.

To begin with, let us review the notion of \emph{totally positive of order 2} ($TP_{2}$). A function $h(x,y)$ is said to be $TP_{2}$ in $(x,y)$, if $h(x,y)\geq0$ and $h(x_{1},y_{1})h(x_{2},y_{2})\geq h(x_{1},y_{2})h(x_{2},y_{1})$, whenever $x_{1}\leq x_{2}$ and $y_{1}\leq y_{2}$. Interested readers are referred to Karlin and Rinott \cite{karlin80} for a comprehensive study on the properties and applications of $TP_{2}$.
\subsection{Series system}
\begin{The}\label{alloicxmin} Suppose the redundancies lifetimes $Y_{1},\ldots,Y_{m}$ have common log-concave density functions. For any $1\leq i<j\leq n$,
\begin{itemize}
\item [(i)] if $\bm{X}$ is LWSAI, then $S(\bm{X}+\bm{Y};\bm{r}) \geq_{\rm{icv}}S(\bm{X}+\bm{Y};\tau_{i,j}(\bm{r}))$ whenever $r_{i}\geq r_{j}$;
\item [(ii)] if $\bm{X}$ is RWSAI, then $S(\bm{X}+\bm{Y};\bm{r}) \geq_{\rm{st}}S(\bm{X}+\bm{Y};\tau_{i,j}(\bm{r}))$ whenever $r_{i}\geq r_{j}$.
\end{itemize}
\end{The}
\proof Define {$Z_{l}=Y_{\sum_{i=1}^{l-1}r_{i}+1}+\cdots+Y_{\sum_{i=1}^{l}r_{i}}$}, for $l=1,2,\ldots,n$, where $r_{0}\equiv0$. Let $f^{(r_{l})}(z_{l})$ be the the density function of $Z_{l}$, which are convolutions of $r_{l}$ copies of $Y$ for $l=1,2,\ldots,n$. For any integrable function $u$, we have
\begin{eqnarray}\label{purpe1}
&&\mathbb{E}[u(S(\bm{X}+\bm{Y};\bm{r}))]-\mathbb{E}[u(S(\bm{X}+\bm{Y};\tau_{i,j}(\bm{r})))] \nonumber\\
 &=& \mathbb{E}[u(\min(\bm{X}+\bm{Z}))]-\mathbb{E}[u(\min(\bm{X}+\tau_{i,j}(\bm{Z})))] \nonumber\\
 &=&\idotsint\limits_{\mathbb{R}_{+}^{n}}\mathbb{E}[u(\min(\bm{X}+\bm{z}))]f^{(r_{i})}(z_{i})f^{(r_{j})}(z_{j})\prod_{l\neq i,j}^{n} f^{(r_{l})}(z_{l}) \dif{z_{1}}\cdots \dif{z_{n}} \nonumber\\
  && -\idotsint\limits_{\mathbb{R}_{+}^{n}}\mathbb{E}[u(\min(\bm{X}+\bm{z}))]f^{(r_{i})}(z_{j})f^{(r_{j})}(z_{i})\prod_{l\neq i,j}^{n} f^{(r_{l})}(z_{l})\dif{z_{1}}\cdots \dif{z_{n}} \nonumber\\
 &=&\idotsint\limits_{\mathbb{R}_{+}^{n-2}}\prod_{l\neq i,j} ^{n}f^{(r_{l})}(z_{l}) \dif z_{l}
 \iint\limits_{z_{i}\leq z_{j}}\bigg\{\big[\mathbb{E}[u(\min(\bm{X}+\bm{z}))]-\mathbb{E}[u(\min(\bm{X}+\tau_{i,j}(\bm{z})))]\big]\nonumber\\
 && \times[f^{(r_{i})}(z_{i})f^{(r_{j})}(z_{j})-f^{(r_{i})}(z_{j})f^{(r_{j})}(z_{i})]\bigg\} \dif{z_{i}} \dif{z_{j}}.
\end{eqnarray}
The desired result is equivalent to proving that (\ref{purpe1}) is non-negative.

On the one hand, according to the proof of Theorem \ref{heterseries} we have shown under cases (i) and (ii) that
\begin{equation}\label{cdieq25}
\mathbb{E}[u(\min(\bm{X}+\bm{z}))]\leq\mathbb{E}[u(\min(\bm{X}+\tau_{i,j}(\bm{z})))],\quad z_{i}\leq z_{j}.
\end{equation}
On the other hand, from the log-concavity of the density function of $Y$, we can conclude that $f^{(r_{l})}(y)$ is $TP_{2}$ in $(r_{l},y)\in\{1,2,\ldots,m\}\times\mathbb{R}_{+}$, where
\begin{equation*}
f^{(r_{l})}(y)=f_{1}(y)\ast f_{2}(y)\ast \cdots \ast f_{r_{l}}(y)
\end{equation*}
denotes the density function of convolutions of $r_{l}$ copies of $Y$, $l=1,2,\ldots,n$. Thus, it can be obtained that
\begin{equation}\label{conve1}
f^{(r_{i})}(z_{i})f^{(r_{j})}(z_{j})-f^{(r_{i})}(z_{j})f^{(r_{j})}(z_{i})\leq0,\quad z_{i}\leq z_{j}~{\rm and}~r_{i}\geq r_{j}.
\end{equation}
Upon combining (\ref{cdieq25}) with (\ref{conve1}), the non-negativity of (\ref{purpe1}) is established. Hence, the theorem follows.\qed

\medskip

\begin{figure}[htp!]
  \centering
  \subfigure[$u(x)=x^{0.8}$]{
    \label{figminhoa} 
    \includegraphics[width=0.48\textwidth,height=0.25\textheight]{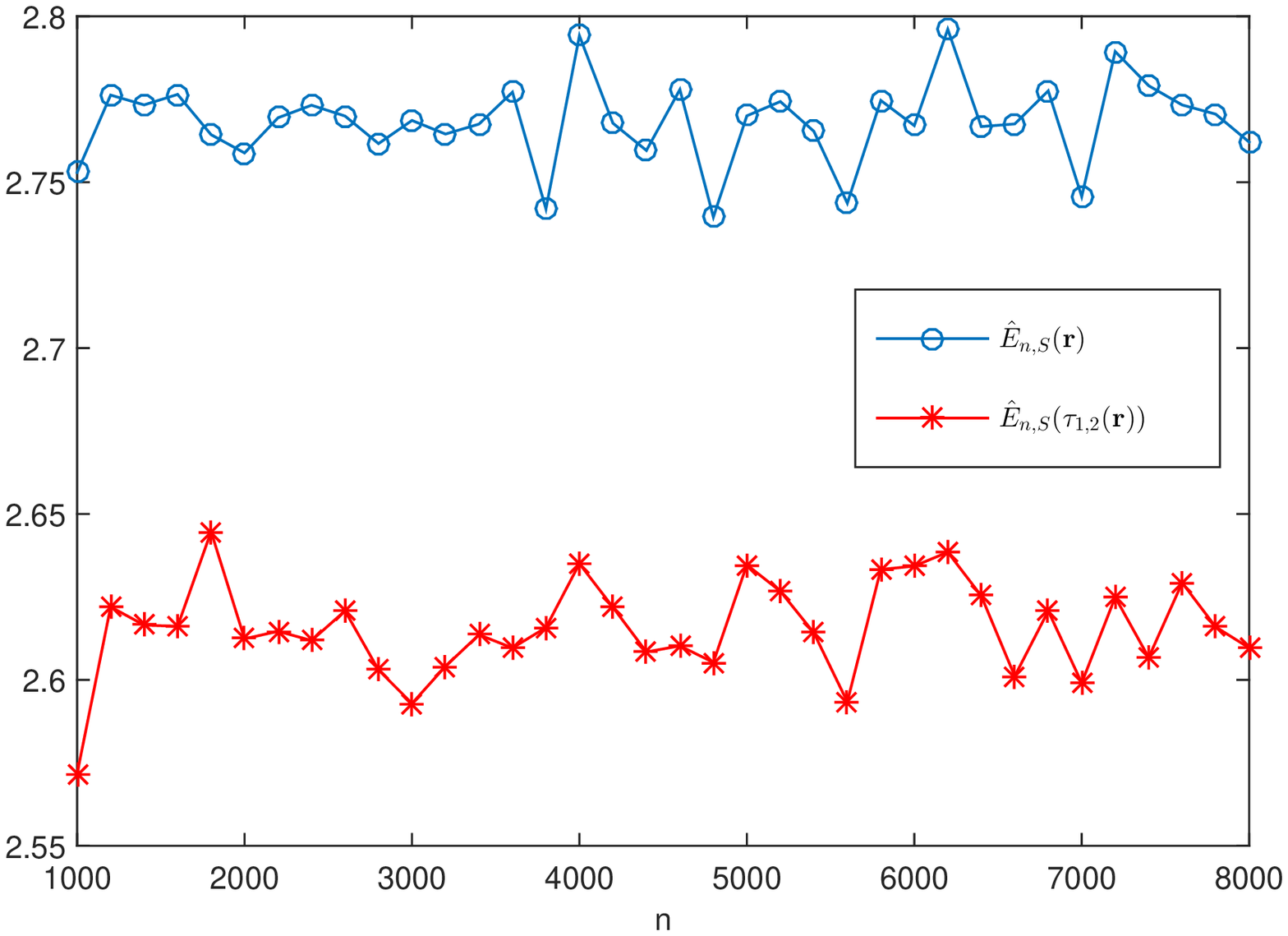}}
  \subfigure[$u(x)=x^{1.2}$]{
    \label{figminhob} 
    \includegraphics[width=0.48\textwidth,height=0.25\textheight]{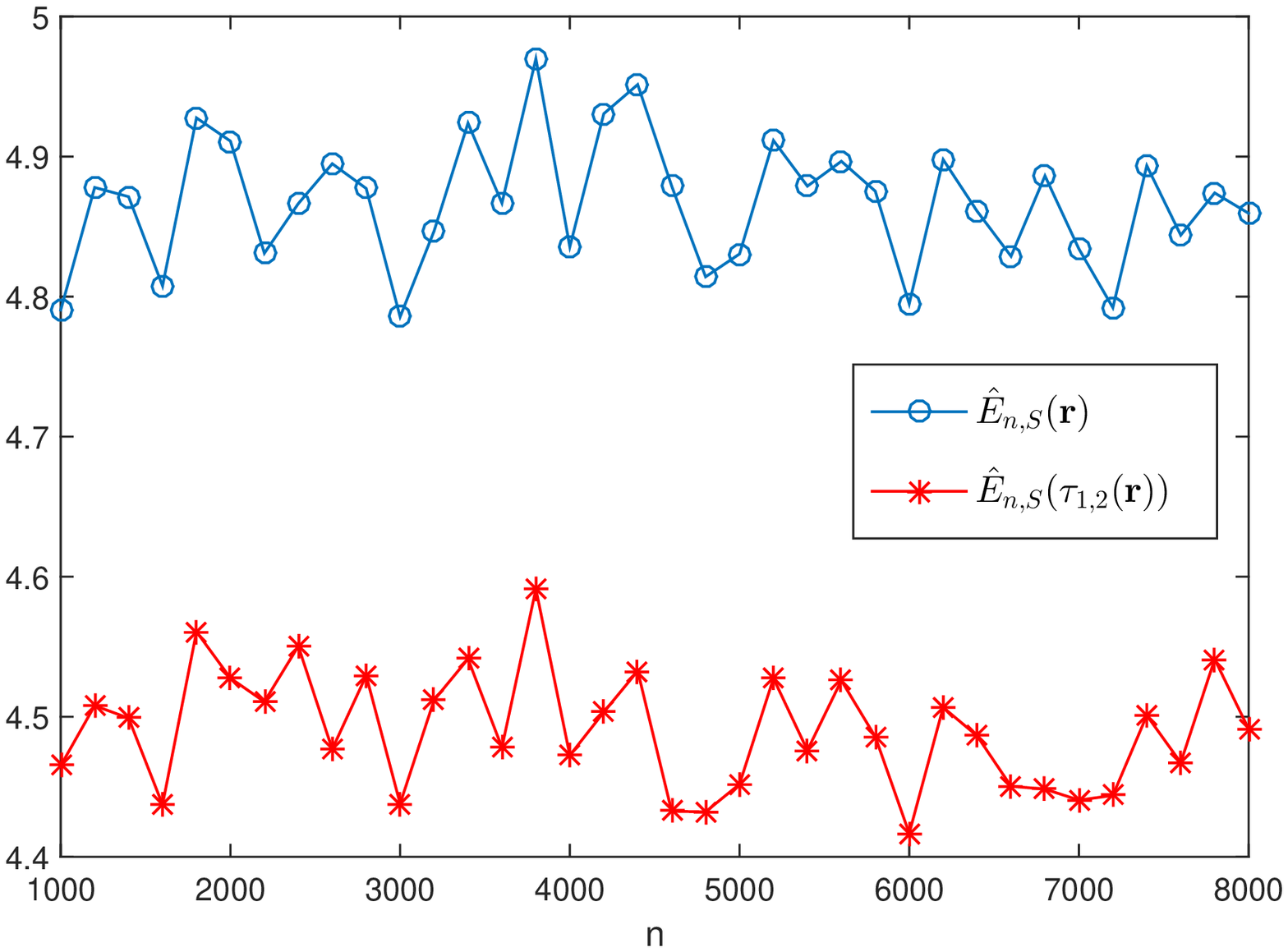}}
    \subfigure[$u(x)=(1-e^{-2x})/2$]{
    \label{figminhoc} 
    \includegraphics[width=0.48\textwidth,height=0.25\textheight]{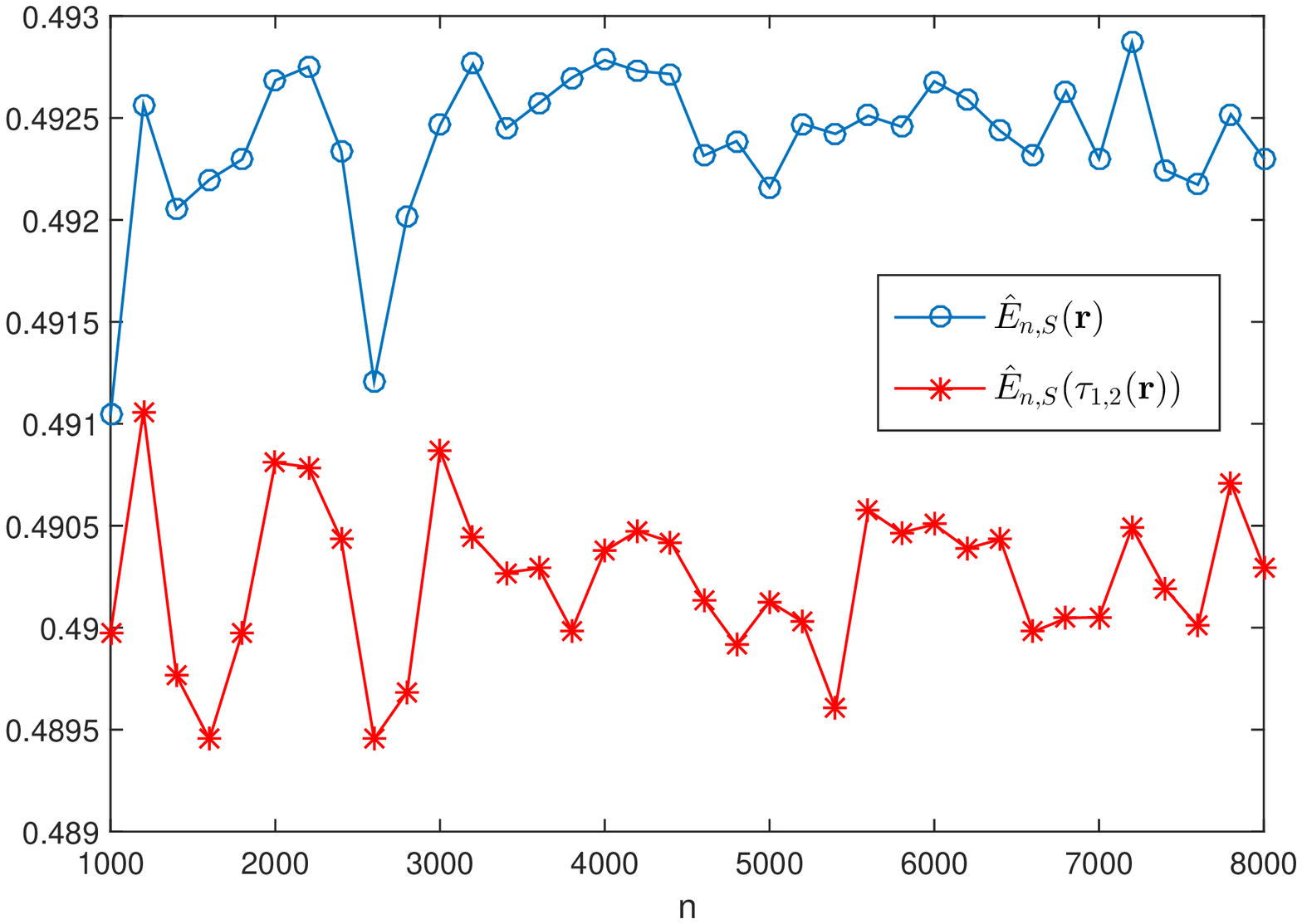}}
    \subfigure[$u(x)=\log x$]{
    \label{figminhoc} 
    \includegraphics[width=0.48\textwidth,height=0.25\textheight]{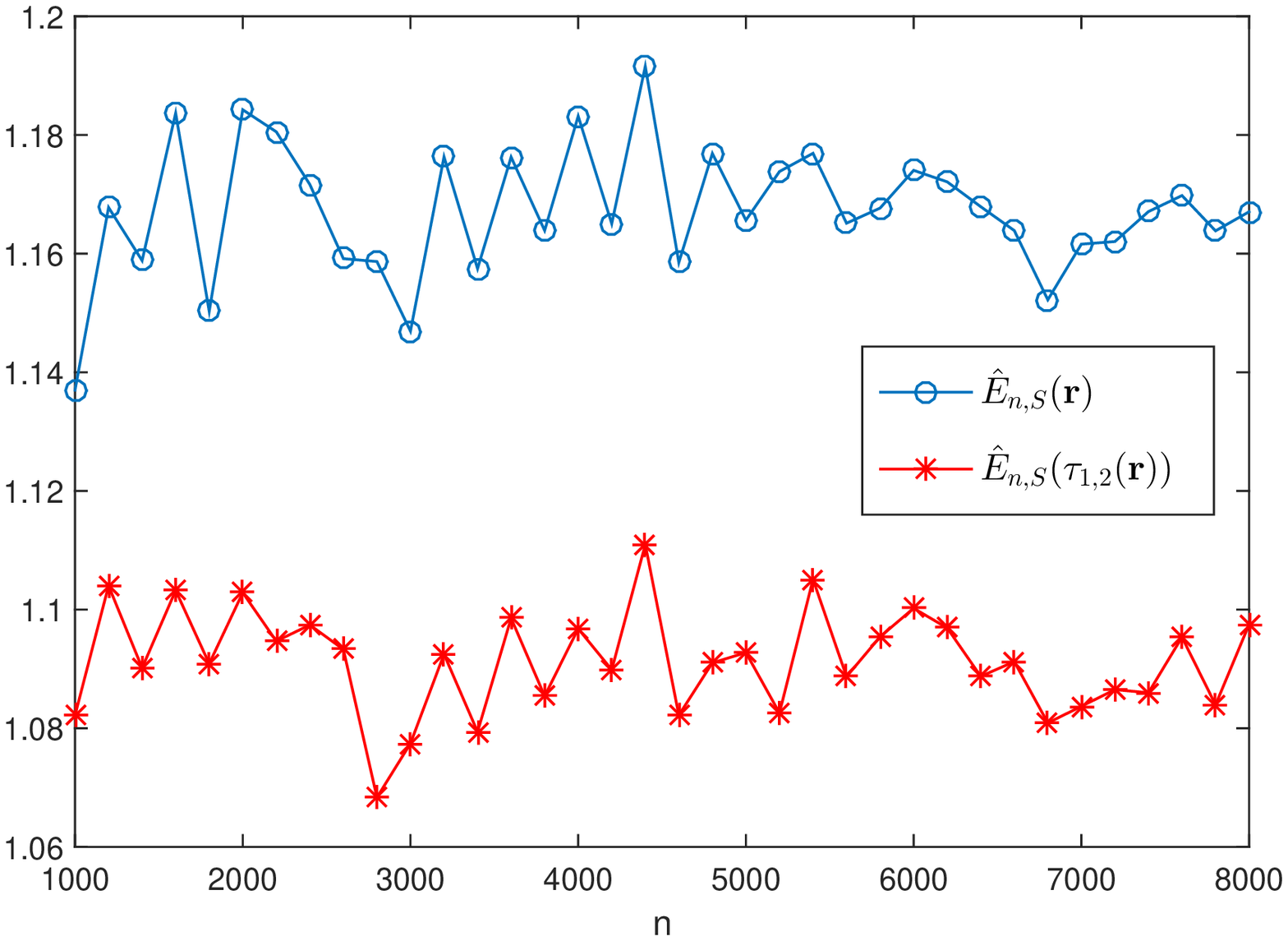}}
  \caption{Plots of the estimators $\widehat{E}_{n,S}(\bm{r})$ and $\widehat{E}_{n,S}(\tau_{1,2}(\bm{r}))$.}
  \label{figminho} 
\end{figure}

Many lifetime distributions have log-concave densities, to name a few, the Beta distribution with both parameters greater than 1, the Gamma distribution with shape parameter greater than 1 and scale parameter equaling to 1, the Weibull distribution with shape parameter greater than 1 and scale parameter equaling to 1, and so on. Based on Theorem \ref{alloicxmin}, it can be figured out that more redundancies should be allocated to the weaker components in the sense of the increasing concave [usual stochastic] ordering when the components lifetimes are LWSAI [RWSAI]. It might be of great interest to investigate whether the log-concavity property of the density for the cold standbys is a must or not in Theorem \ref{alloicxmin}. Numerical examples show that this assumption might be relaxed or discarded, for which we cannot prove so far and thus leave it as an open problem.

The following numerical example shows the validity of Theorem \ref{alloicxmin}.
\begin{Exa}\label{examinhomo}
Assume $\bm{X}=(X_{1},X_{2})$ is assembled with Clayton [survival] copula with generator $\phi(t)=(t+1)^{-1}$ and $X_{1}$ and $X_{2}$ have exponential distribution with hazard rates $\lambda_{1}=0.5$ and $\lambda_{2}=0.3$, respectively. Let $Y_{1},\ldots,Y_{5}$ be independent Weibull random variables with common scale parameter $\mu=1$ and shape parameter $\beta=1.3$. It is easy to check that $\bm{X}$ is LWSAI [RWSAI], and $Y_{1}$ has log-concave density function. Let $\bm{r}=(3,2)$. For any increasing function $u$, we denote
\begin{equation*}
E_{u,S}(\bm{r}):=\mathbb{E}[u(\min\{X_{1}+Y_{1}+Y_{2}+Y_{3},X_{2}+Y_{4}+Y_{5}\})]
\end{equation*}
and
\begin{equation*}
E_{u,S}(\tau_{1,2}(\bm{r})):=\mathbb{E}[u(\min\{X_{1}+Y_{1}+Y_{2},X_{2}+Y_{3}+Y_{4}+Y_{5}\})].
\end{equation*}
From the population $(X_{1},X_{2},Y_{1},\ldots,Y_{5})$, we generate an i.i.d. sample
\begin{equation*}
(X_{1,1},X_{2,1},Y_{1,1},\ldots,Y_{5,1}),\ldots,(X_{1,n},X_{2,n},Y_{1,n},\ldots,Y_{5,n}),
\end{equation*}
{where the samples $(X_{1,i},X_{2,i})$'s are generated via the method in Subsection 2.9 of Nelsen \cite{nelsen06}.} Then, the functions $E_{u,S}(\bm{r})$ and $E_{u,S}(\tau_{1,2}(\bm{r}))$ can be approximated by
\begin{equation*}
\widehat{E}_{n,S}(\bm{r})=\frac{1}{n}\sum_{i=1}^{n}u(\min\{X_{1,i}+Y_{1,i}+Y_{2,i}+Y_{3,i},X_{2,i}+Y_{4,i}+Y_{5,i}\})
\end{equation*}
and
\begin{equation*}
\widehat{E}_{n,S}(\tau_{1,2}(\bm{r}))=\frac{1}{n}\sum_{i=1}^{n}u(\min\{X_{1,i}+Y_{1,i}+Y_{2,i},X_{2,i}+Y_{3,i}+Y_{4,i}+Y_{5,i}\}),
\end{equation*}
respectively. As observed in Figure \ref{figminho}, {for utility functions $u(x)=x^{0.8}$, $u(x)=x^{1.2}$, $u(x)=(1-e^{-2x})/2$ and $u(x)=\log x$}, it holds that $\widehat{E}_{n,S}(\bm{r})\geq\widehat{E}_{n,S}(\tau_{1,2}(\bm{r}))$ for $n=1000,1200,\ldots,8000$. By law of large numbers, we have $E_{u,S}(\bm{r})\geq E_{u,S}(\tau_{1,2}(\bm{r}))$, which supports the result of Theorem \ref{alloicxmin}.
\end{Exa}

{The following example sheds light on the optimal allocation strategies for series systems, which cannot be proven so far due to technicality difficulty and is left as an open problem.
\begin{Exa} Under the setup of Example \ref{examinhomo}, we plot the values of $\widehat{E}_{n,S}(\bm{r})$ in Figure \ref{figminoptim} for three allocation policies $\bm{r}_1=(3,2)$, $\bm{r}_2=(4,1)$, and $\bm{r}_3=(5,0)$. These numerical simulations suggest that the optimal allocation policy for a series system might be such that (i) more {cold-standby} redundancies should be allocated to weaker components, and (ii) the number of spares given in each node should be as close as possible.
\end{Exa}
\begin{figure}[htp!]
	\centering
	\subfigure[$u(x)=x^{0.8}$]{
		\label{figminoptim1} 
		\includegraphics[width=0.48\textwidth,height=0.25\textheight]{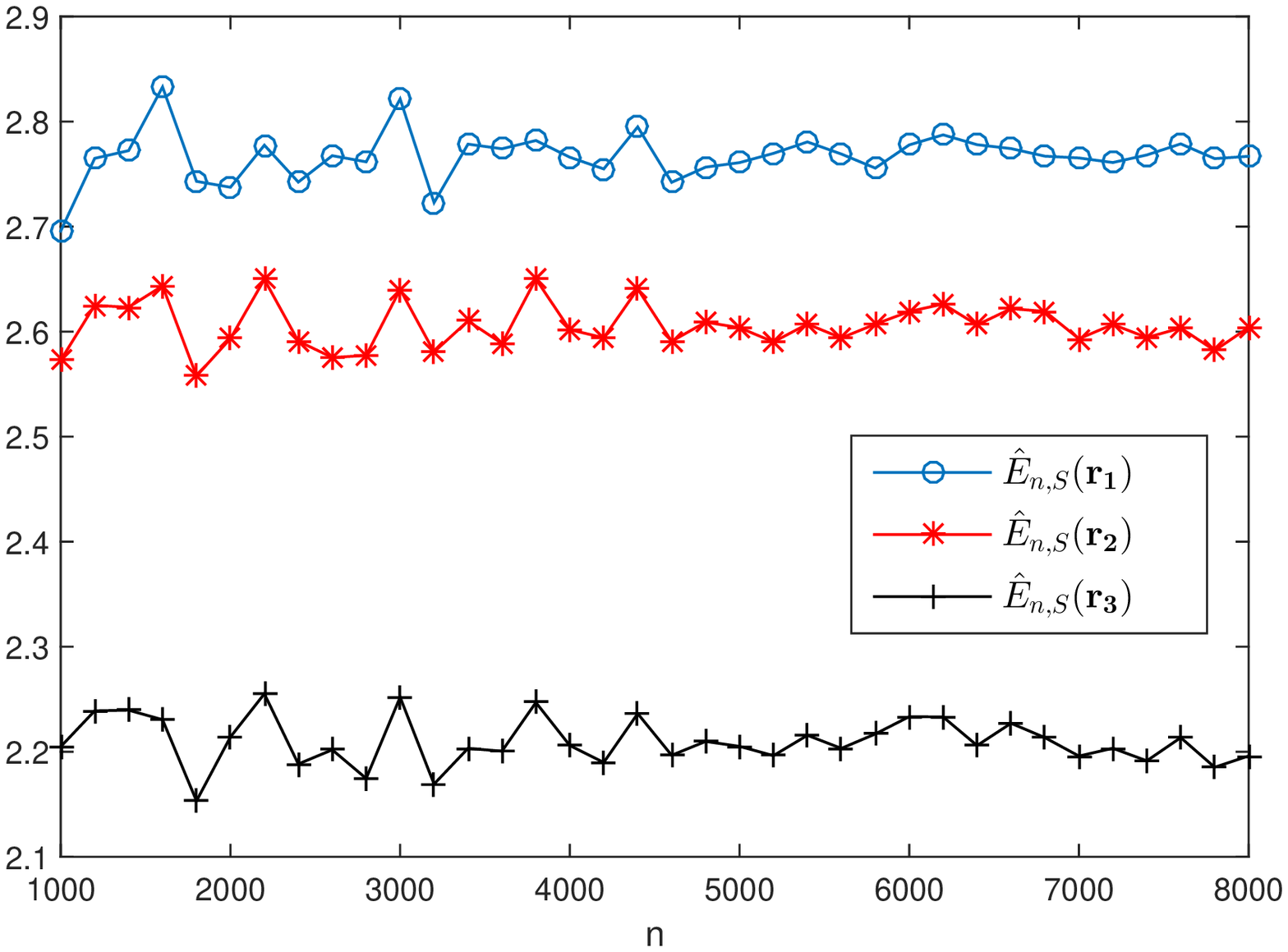}}
	\subfigure[$u(x)=x^{1.2}$]{
		\label{figminoptim2} 
		\includegraphics[width=0.48\textwidth,height=0.25\textheight]{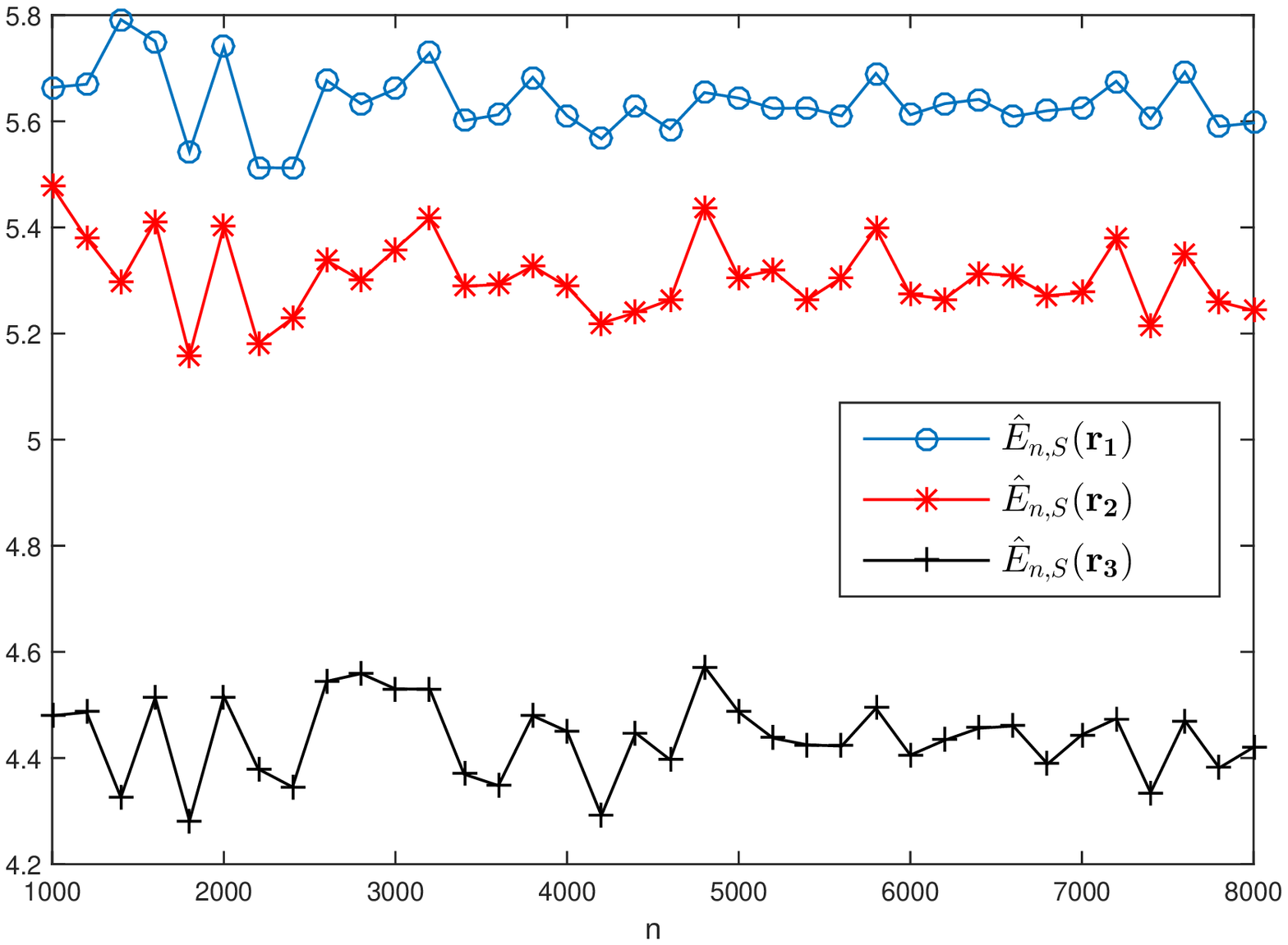}}
	\caption{Plots of the estimators $\widehat{E}_{n,S}(\bm{r})$ for different allocation policies $\bm{r}$.}
	\label{figminoptim} 
\end{figure}}

The following corollary can be obtained from Theorem \ref{alloicxmin}, which partially generalizes Theorem 4.2 of Zhuang and Li \cite{zhuang15} to the case of dependent components.

\begin{Cor}\label{coromin} Suppose the redundancies lifetimes $Y_{1},\ldots,Y_{m}$ and have common log-concave density functions. For any $1\leq i<j\leq n$,
\begin{itemize}
\item [(i)] if $\bm{X}$ has an Archimedean copula with log-convex generator and such that $X_{1}\leq_{\rm rh}\cdots\leq_{\rm rh}X_{n}$, then $S(\bm{X}+\bm{Y};\bm{r}) \geq_{\rm{icv}}S(\bm{X}+\bm{Y};\tau_{i,j}(\bm{r}))$ whenever $r_{i}\geq r_{j}$;
\item [(ii)] if $\bm{X}$ has an Archimedean survival copula with log-convex generator and such that $X_{1}\leq_{\rm hr}\cdots\leq_{\rm hr}X_{n}$, then $S(\bm{X}+\bm{Y};\bm{r}) \geq_{\rm{st}}S(\bm{X}+\bm{Y};\tau_{i,j}(\bm{r}))$ whenever $r_{i}\geq r_{j}$.
\end{itemize}
\end{Cor}

\subsection{Parallel system}
In this subsection, we present optimal allocation strategies of i.i.d. {cold-standby} redundancies to a parallel system comprised of dependent components having LWSAI or RWSAI joint lifetimes.
\begin{The}\label{allomax} Suppose the redundancies lifetimes $Y_{1},\ldots,Y_{m}$ have common log-concave density functions. For any $1\leq i<j\leq n$,
\begin{itemize}
\item [(i)] if $\bm{X}$ is LWSAI, then $T(\bm{X}+\bm{Y};\bm{r}) \leq_{\rm{st}}T(\bm{X}+\bm{Y};\tau_{i,j}(\bm{r}))$ whenever $r_{i}\geq r_{j}$;
\item [(ii)] if $\bm{X}$ is RWSAI, then $T(\bm{X}+\bm{Y};\bm{r}) \leq_{\rm{icx}}T(\bm{X}+\bm{Y};\tau_{i,j}(\bm{r}))$ whenever $r_{i}\geq r_{j}$.
\end{itemize}
\end{The}
\proof In light of the proof of Theorem \ref{alloicxmin}, one can see that
\begin{eqnarray}\label{purpos1}
&&\mathbb{E}[u(T(\bm{X}+\bm{Y};\bm{r}))]-\mathbb{E}[u(T(\bm{X}+\bm{Y};\tau_{i,j}(\bm{r})))] \nonumber\\
 &=& \mathbb{E}[u(\max(\bm{X}+\bm{Z}))]-\mathbb{E}[u(\max(\bm{X}+\tau_{i,j}(\bm{Z})))] \nonumber\\
 &=&\idotsint\limits_{\mathbb{R}_{+}^{n-2}}\prod_{l\neq i,j} ^{n}f^{(r_{l})}(z_{l}) \dif z_{l}
 \iint\limits_{z_{i}\leq z_{j}}\bigg\{\big[\mathbb{E}[u(\max(\bm{X}+\bm{z}))]-\mathbb{E}[u(\max(\bm{X}+\tau_{i,j}(\bm{z})))]\big]\nonumber\\
 && \times[f^{(r_{i})}(z_{i})f^{(r_{j})}(z_{j})-f^{(r_{i})}(z_{j})f^{(r_{j})}(z_{i})]\bigg\} \dif{z_{i}} \dif{z_{j}}.
\end{eqnarray}
Then, the non-positivity of (\ref{purpos1}) can be established under cases (i) and (ii) by using (\ref{conve1}) and the proof method as in Theorem \ref{heterparallel}.\qed

\medskip

It can inferred from Theorem \ref{allomax} that more redundancies should be allocated to better components in order to reach a resulting parallel system with higher lifetime in the sense of the usual stochastic [increasing convex] ordering when the joint lifetimes of the original components are LWSAI [RWSAI]. {It remains open to investigate whether the log-concavity assumption on the common density function of the cold standbys could be relaxed (or removed) or not.}

The following example illustrates Theorem \ref{allomax}.
\begin{Exa}\label{exaparax}
\begin{figure}[htp!]
  \centering
  \subfigure[$u(x)=x^{0.8}$]{
    \label{figmaxhoa} 
    \includegraphics[width=0.48\textwidth,height=0.25\textheight]{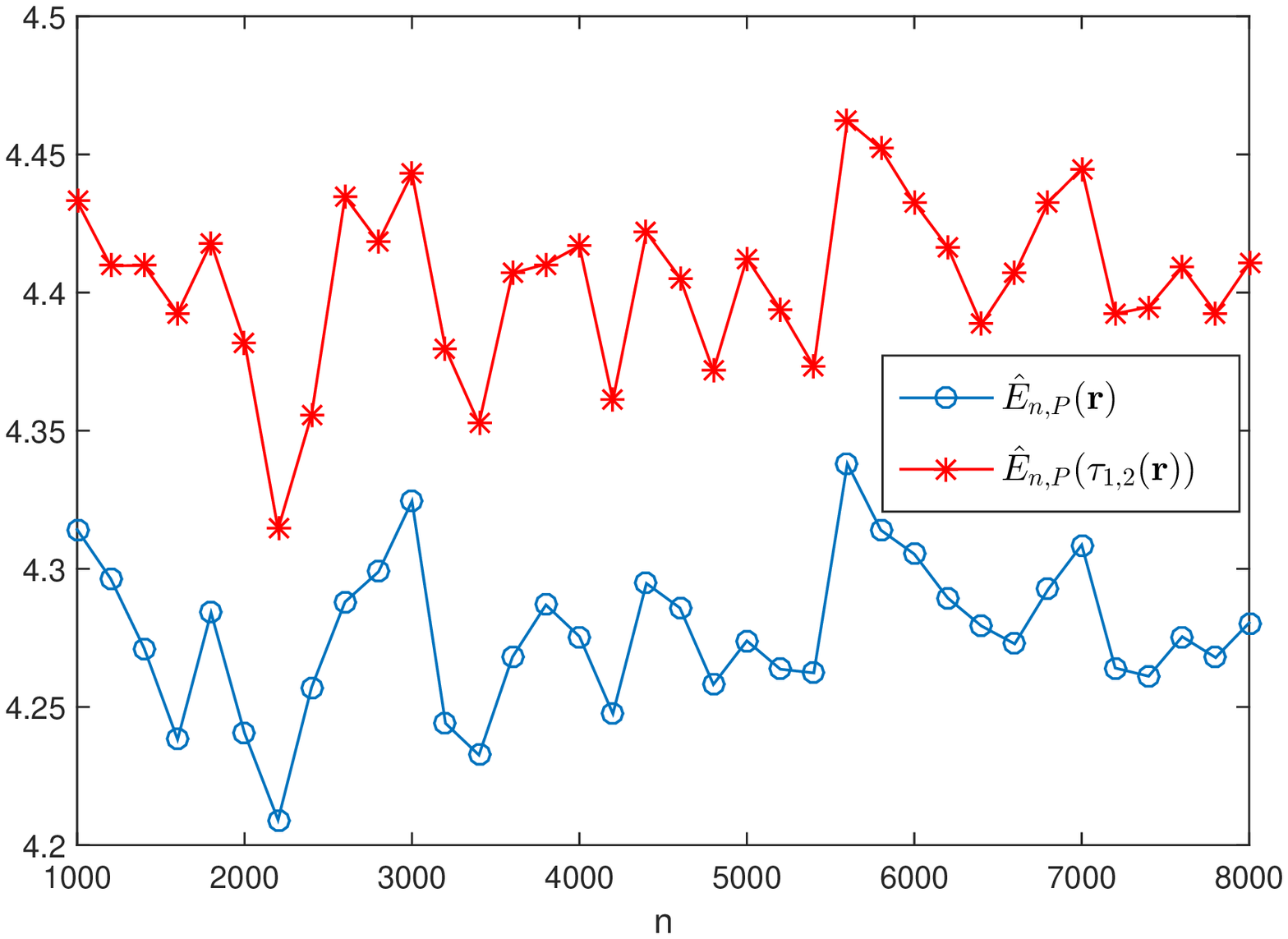}}
  \subfigure[$u(x)=x^{1.2}$]{
    \label{figmaxhob} 
    \includegraphics[width=0.48\textwidth,height=0.25\textheight]{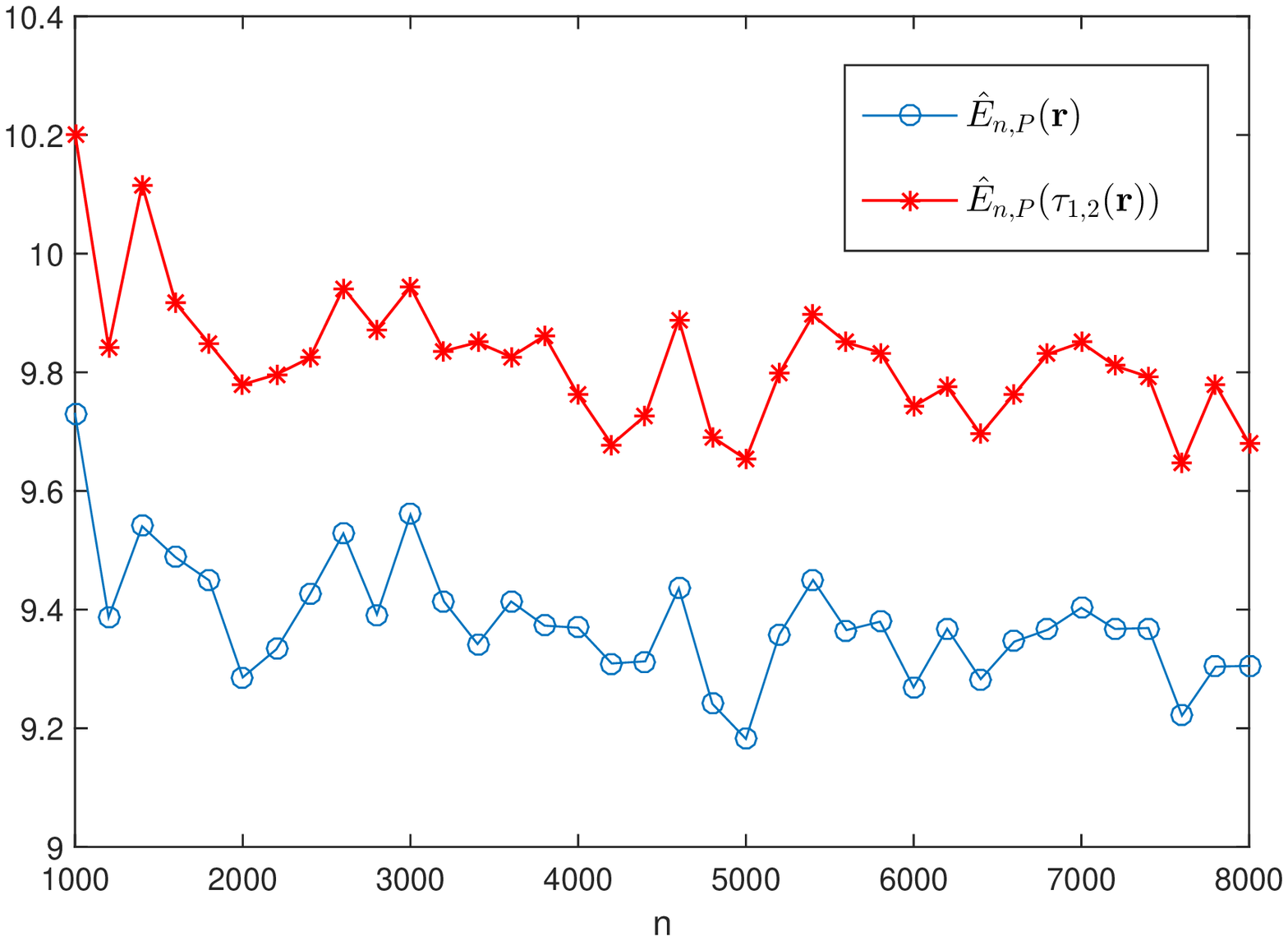}}
    \subfigure[$u(x)=10(1-e^{-0.1x})$]{
    \label{figmaxhoc} 
    \includegraphics[width=0.48\textwidth,height=0.25\textheight]{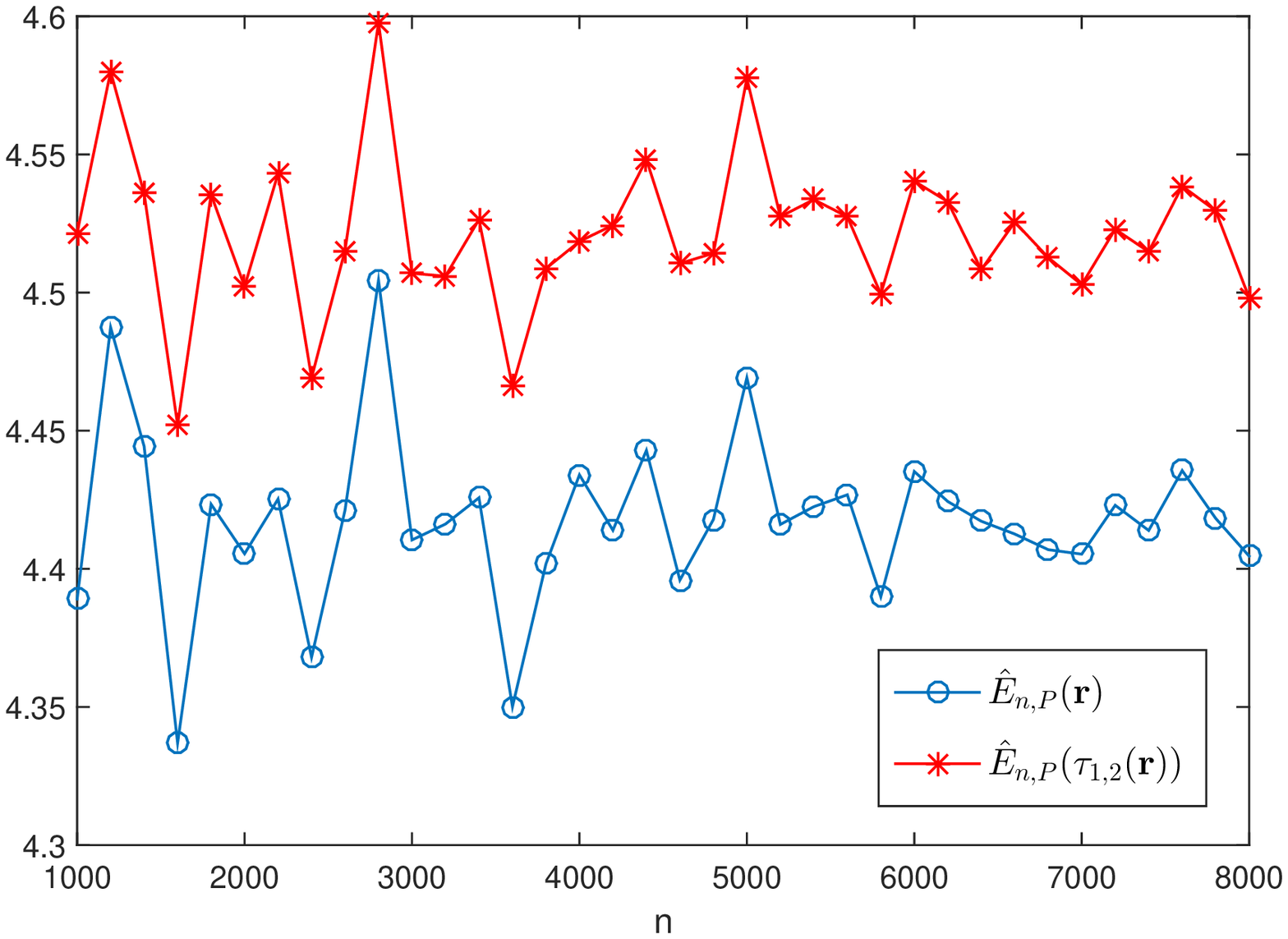}}
    \subfigure[$u(x)=\log x$]{
    \label{figmaxhod} 
    \includegraphics[width=0.48\textwidth,height=0.25\textheight]{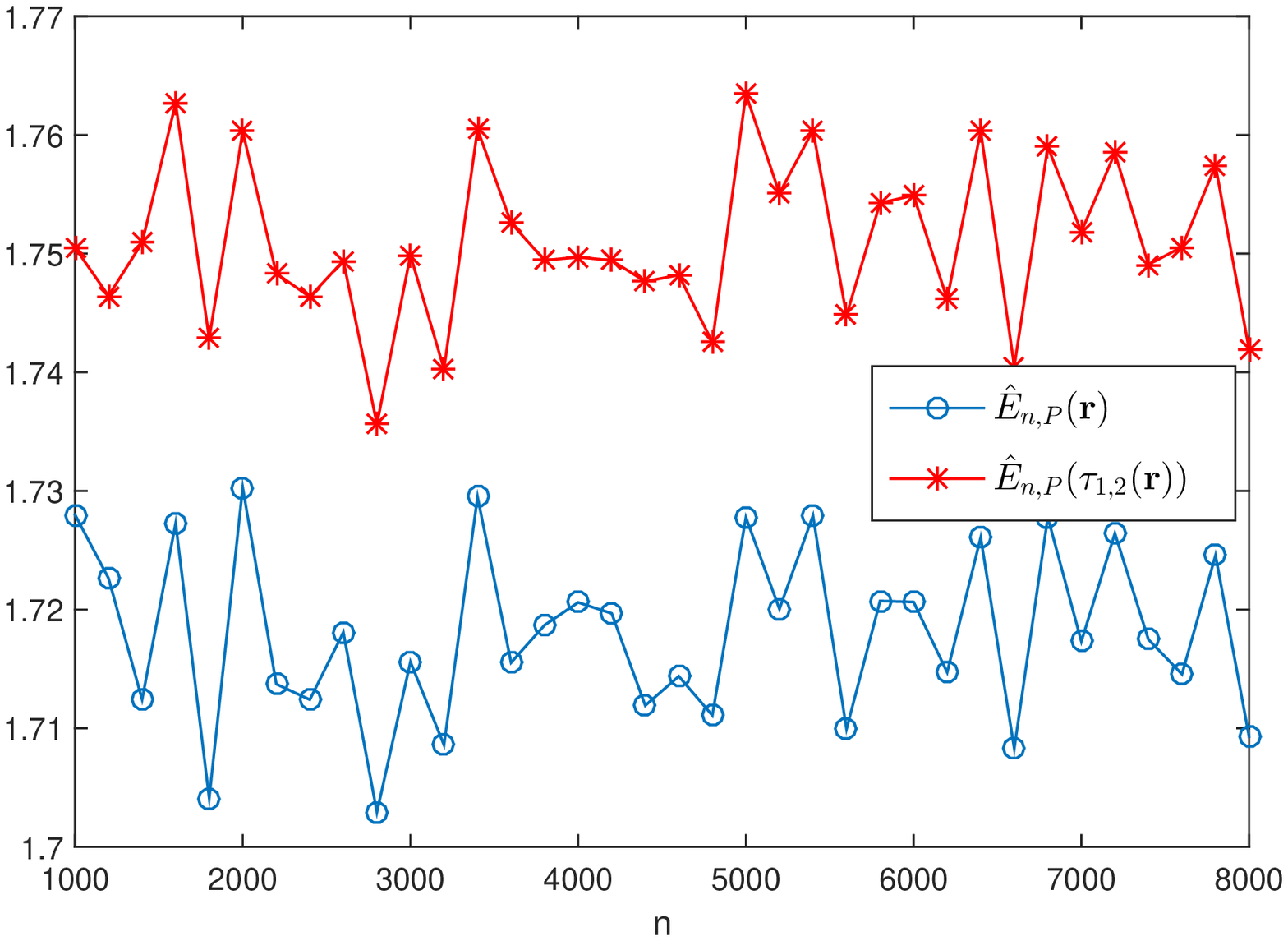}}
  \caption{Plots of the estimators $\widehat{E}_{n,P}(\bm{r})$ and $\widehat{E}_{n,P}(\tau_{1,2}(\bm{r}))$.}
  \label{figmaxho} 
\end{figure}
Under the setup of Example \ref{examinhomo}, for any increasing function $u$, we denote
\begin{equation*}
E_{u,P}(\bm{r}):=\mathbb{E}[u(\max\{X_{1}+Y_{1}+Y_{2}+Y_{3},X_{2}+Y_{4}+Y_{5}\})]
\end{equation*}
and
\begin{equation*}
E_{u,P}(\tau_{1,2}(\bm{r})):=\mathbb{E}[u(\max\{X_{1}+Y_{1}+Y_{2},X_{2}+Y_{3}+Y_{4}+Y_{5}\})].
\end{equation*}
The function $E_{u,P}(\bm{r})$ and $E_{u,P}(\tau_{1,2}(\bm{r}))$ can be approximated by
\begin{equation*}
\widehat{E}_{n,P}(\bm{r})=\frac{1}{n}\sum_{i=1}^{n}u(\max\{X_{1,i}+Y_{1,i}+Y_{2,i}+Y_{3,i},X_{2,i}+Y_{4,i}+Y_{5,i}\})
\end{equation*}
and
\begin{equation*}
\widehat{E}_{n,P}(\tau_{1,2}(\bm{r}))=\frac{1}{n}\sum_{i=1}^{n}u(\max\{X_{1,i}+Y_{1,i}+Y_{2,i},X_{2,i}+Y_{3,i}+Y_{4,i}+Y_{5,i}\}),
\end{equation*}
respectively. Figure \ref{figmaxho} plots the estimators $\widehat{E}_{n,P}(\bm{r})$ and $\widehat{E}_{n,P}(\tau_{1,2}(\bm{r}))$ for $n=1000,1200,\ldots,8000$ under utility functions {$u(x)=x^{0.8}$, $u(x)=x^{1.2}$, $u(x)=10(1-e^{-0.1x})$, and $u(x)=\log x$}. Then, by law of large numbers it holds that $E_{u,P}(\bm{r})\leq E_{u,P}(\tau_{1,2}(\bm{r}))$, which validates Theorem \ref{allomax}.
\end{Exa}

\begin{figure}[htp!]
	\centering
	\subfigure[$\alpha=0.8$]{
		\label{figmaxoptim1} 
		\includegraphics[width=0.48\textwidth,height=0.25\textheight]{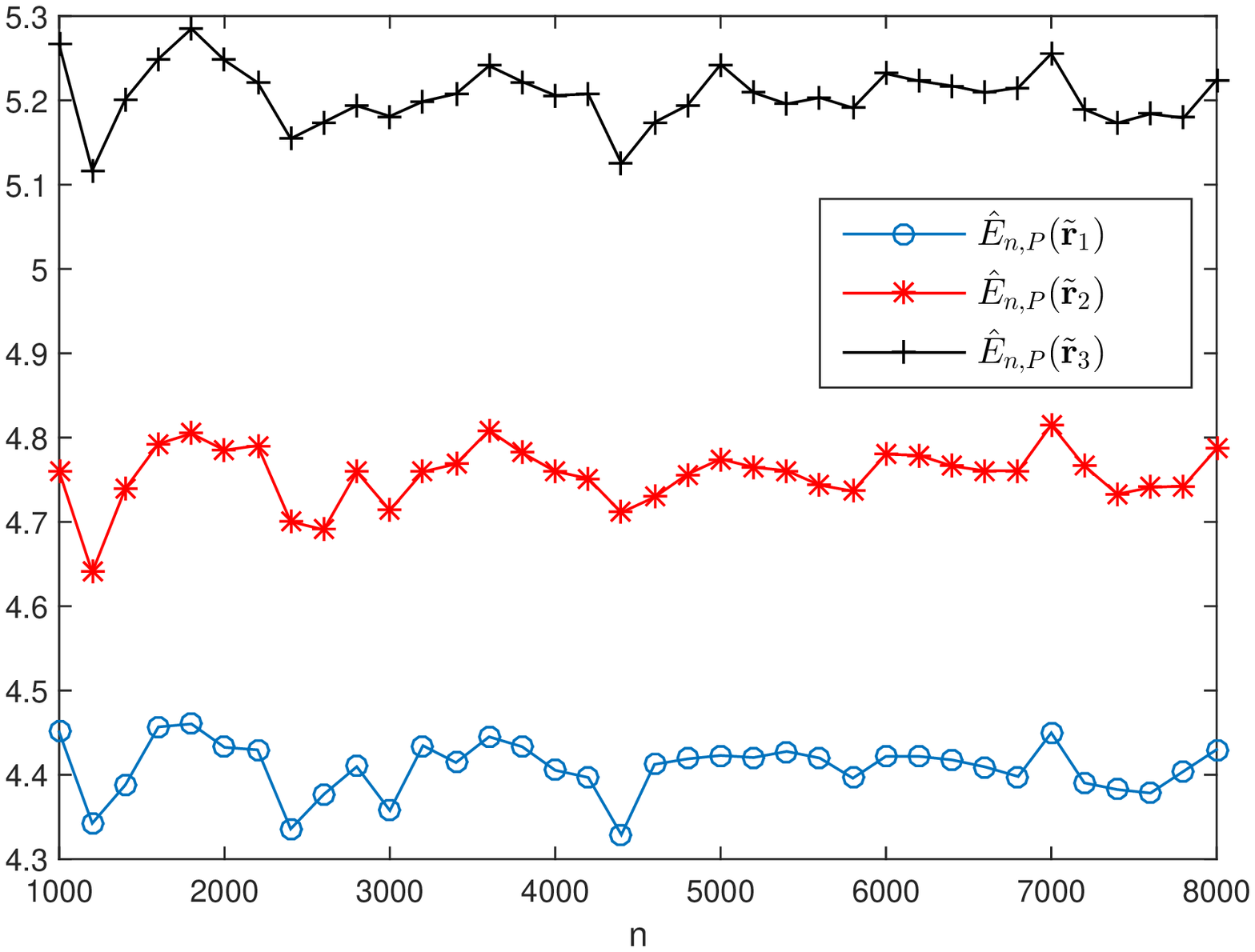}}
	\subfigure[$\alpha=1.2$]{
		\label{figmaxoptim2} 
		\includegraphics[width=0.48\textwidth,height=0.25\textheight]{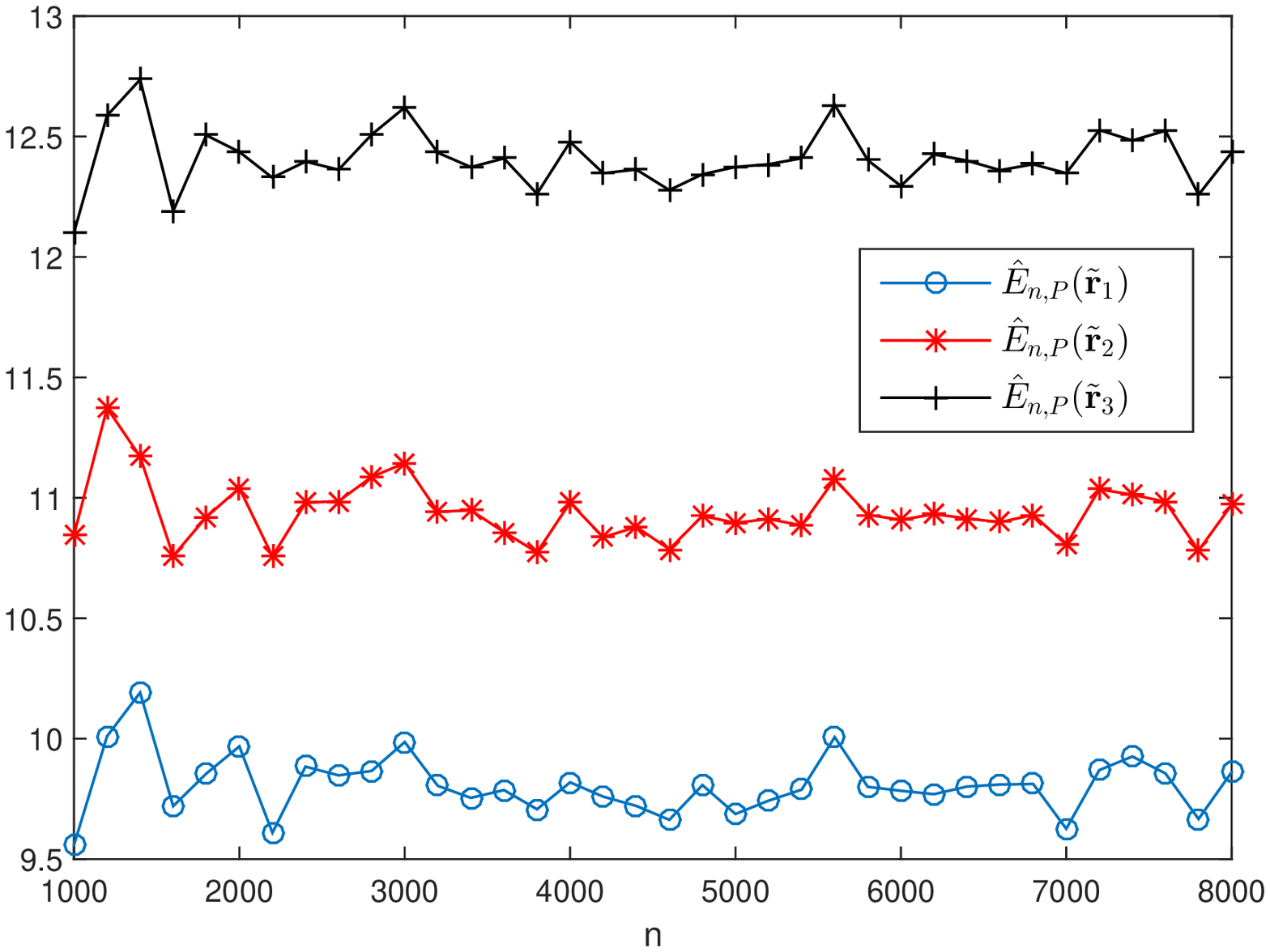}}
	\caption{Plots of the estimators $\widehat{E}_{n,P}(\tilde{\bm{r}})$ for different allocation policies $\tilde{\bm{r}}$.}
	\label{figmaxoptim} 
\end{figure}

{However, the explicit configuration of the optimal allocation strategy remains to be determined for parallel systems. The following example conjectures on the optimal allocation, which cannot be proven so far due to technical difficulty and is thus left as an open problem.
\begin{Exa} The values of $\widehat{E}_{n,P}(\tilde{\bm{r}})$ are displayed in Figure \ref{figmaxoptim} for three policies $\tilde{\bm{r}}_1=(2,3)$, $\tilde{\bm{r}}_2=(1,4)$ and $\tilde{\bm{r}}_3=(0,5)$ under the setting of Example \ref{exaparax}. These plots suggest that, for a parallel system, all redundancies might be allocated to the component with the best performance, which agrees with intuition.
\end{Exa}}

Similar with Corollary \ref{coromin}, the following result can be obtained from Theorem \ref{allomax}.
\begin{Cor}\label{coromax} Suppose the redundancies lifetimes $Y_{1},\ldots,Y_{m}$ have common log-concave density functions. For any $1\leq i<j\leq n$,
\begin{itemize}
\item [(i)] if $\bm{X}$ has an Archimedean copula with log-convex generator and such that $X_{1}\leq_{\rm rh}\cdots\leq_{\rm rh}X_{n}$, then $T(\bm{X}+\bm{Y};\bm{r}) \geq_{\rm{st}}T(\bm{X}+\bm{Y};\tau_{i,j}(\bm{r}))$ whenever $r_{i}\geq r_{j}$;
\item [(ii)] if $\bm{X}$ has an Archimedean survival copula with log-convex generator and such that $X_{1}\leq_{\rm hr}\cdots\leq_{\rm hr}X_{n}$, then $T(\bm{X}+\bm{Y};\bm{r}) \geq_{\rm{icx}}T(\bm{X}+\bm{Y};\tau_{i,j}(\bm{r}))$ whenever $r_{i}\geq r_{j}$.
\end{itemize}
\end{Cor}

\section{Conclusions}\label{seccon}
In reliability theory and engineering practice, it is an important research issue to seek for optimal redundancies allocation strategies for coherent systems. In this article, we investigate optimal allocation policies of {cold-standby} redundancies in series and parallel systems comprised of dependent components having LWSAI or RWSAI joint lifetimes. {Under the assumption that the {cold-standby} redundancies are independent of the original components,} optimal allocations are pinpointed both for series and parallel systems under matching allocation when the {cold-standby} spares are independent and ordered via the likelihood ratio ordering. For the homogeneous {cold-standby} redundancies, optimal allocation policies are also derived for both series and parallel systems. The optimal allocation strategies for series systems are opposite to those for parallel systems, which are consistent with the findings in Boland et al. \cite{boland92}.

You et al. \cite{you16} established the optimal allocations of hot-standby redundancies for $k$-out-of-$n$ systems with components having LTPD lifetimes. They applied the allocation strategies in distributing new wires to a set of cables in order to increase the strength of the cables. Cables with great tensile strength are commonly demanded for designing a high voltage electricity transmission network. The strength of a cable composed of several wires can be viewed as a $k$-out-of-$n$ system, where the wires may be regarded as components. Consider the selected wires labeled as 1, 2 and 8 in the data set tested and reported in Hald \cite{Hald52} and embodied in Hand et al. \cite{Hand94}. Denote the tensile strength of wires 1, 2, 8 by $X_1$, $X_2$ and $X_8$, respectively. By showing that Fr\'echet distribution fits the strength of these wires well, and that the dependence structure is fitted by Gumbel-Barnett copula statistically well, You et al. \cite{you16} shows that the vector $(X_2,X_1,X_8)$ can be modelled by some absolutely continuous LTPD distribution, which is equivalent to LSWAI. Therefore, if these three wires are assembled in series or parallel, the allocation strategies derived in Sections \ref{sechetero} and \ref{sechomo} can be applied in the cable to increase the strength. For example, assume that a cable is made from these three wires, and its strength is measured as the weakest of the wires. Consider the repairing strategy such that a broken wire is replaced by a new one. Ignoring the replacing time, the lifetime of the cable can be approximated by that of a series system under some cold-standby allocation strategy. If there are three extra wires with independent strengths ordered in the sense of the likelihood ratio order, each of which is to replace one broken wire, then the best replacing strategy is to put the strongest one to wire 2, the moderate one to wire 1, and the weakest one to wire 8.

{Since both LWSAI and RWSAI are positive dependence notions, it is of natural interest to study whether the optimal allocation policies established here still hold for the case of negatively dependent components. Besides, another possible extension might be seeking for the best allocation policies when the original components have WSAI (c.f. Cai and Wei, \cite{caiwei15}) joint lifetimes. We leave these as open problems.}

\section*{Acknowledgments}
The authors are very grateful for the valuable comments from two anonymous reviewers, which have greatly improved the presentation of this paper. Yiying Zhang thanks the start-up grant at Nankai University.

\end{document}